\documentclass[fleqn,usenatbib]{mnras}

\usepackage{newtxtext,newtxmath}

\usepackage[T1]{fontenc}
\usepackage{tabularx}

\DeclareRobustCommand{\VAN}[3]{#2}
\let\VANthebibliography\thebibliography
\def\thebibliography{\DeclareRobustCommand{\VAN}[3]{##3}\VANthebibliography}

\usepackage{graphicx}	
\usepackage{amsmath}	
\usepackage{orcidlink}
\usepackage{xcolor}
\usepackage[percent]{overpic}

\definecolor{red}{rgb}{0.96, 0.36, 0.36}

\title[MW--LMC Interaction with SBI]{The Milky Way -- Large Magellanic Cloud Interaction with Simulation Based Inference}
\author[R. A. N. Brooks et al.]{Richard A. N. Brooks$^{1,2}$\thanks{E-mail: richard.brooks.22@ucl.ac.uk}\orcidlink{0000-0001-5550-2057},
Jason L. Sanders$^{1}$\orcidlink{0000-0003-4593-6788},
Vedant Chandra$^{3}$\orcidlink{0000-0002-0572-8012},
Nicolás Garavito-Camargo$^{4}$\thanks{NASA Hubble Fellowship Program, Einstein Fellow}\orcidlink{0000-0001-7107-1744}, \newauthor
Adam M. Dillamore$^{1}$\orcidlink{0000-0003-0807-5261},
Adrian~M.~Price-Whelan$^{2}$\orcidlink{0000-0003-0872-7098}, 
Yuan-Sen Ting$^{5,6}$\orcidlink{000-0001-5082-9536}
\\
$^{1}$Department of Physics and Astronomy, University College London, London, WC1E 6BT, UK\\
$^{2}$Center for Computational Astrophysics, Flatiron Institute, Simons Foundation, 162 Fifth Avenue, New York, NY 10010, USA\\
$^{3}$ Center for Astrophysics | Harvard \& Smithsonian, 60 Garden Street, Cambridge, MA 02138, USA\\
$^{4}$ Department of Astronomy, University of Maryland, College Park, MD 20742, USA\\
$^{5}$ Department of Astronomy, The Ohio State University, Columbus, OH 43210, USA\\
$^{6}$ Center for Cosmology and AstroParticle Physics (CCAPP), The Ohio State University, Columbus, OH 43210, USA\\
}

\date{Accepted XXX. Received YYY; in original form ZZZ}

\pubyear{\the\year{}}

\begin{document}
\label{firstpage}
\pagerange{\pageref{firstpage}--\pageref{lastpage}}
\maketitle

\begin{abstract}
The infall of the Large Magellanic Cloud (LMC) into the Milky Way (MW) has displaced the MW's centre of mass, manifesting as an observed reflex motion in the velocities of outer halo stars. 
We use a Simulation Based Inference framework to constrain properties of the MW, LMC and the induced reflex motion using the dynamics of outer MW halo stars. 
Specifically, we use the mean radial and tangential velocities of outer halo stars calculated in a set of distance and on-sky bins. 
We train neural networks to estimate parameter posterior distributions using a set of $128,000$ rigid MW--LMC simulations conditioned upon velocity data from the Dark Energy Spectroscopic Instrument (DESI) and the combined H3+SEGUE+MagE outer halo surveys. 
We constrain the reflex motion velocity and the enclosed LMC mass within $50 \, \rm kpc$ using the DESI or H3+SEGUE+MagE dataset while varying the survey sky coverage and depth.
Using the radial and tangential velocity data from the H3+SEGUE+MagE survey and on-sky quadrants, we report a distance-averaged reflex motion velocity for the outer halo samples, the speed at which the MW lurches towards the LMC, of $v_{\rm{travel}} = 26.4^{+5.5}_{-4.4} \, \rm km \, \rm s^{-1}$, while simultaneously finding an enclosed LMC mass of $M_{\rm LMC}(< 50 \, \rm kpc) = 9.2^{+1.9}_{-2.3} \times 10^{10}\, \rm M_{\odot}$. Quoted uncertainties are statistical.
Our results suggest that the LMC's total mass is at least $\approx 10-15 \%$ of that of the MW.
This inference framework is flexible such that it can provide rapid constraints when applied to any future survey measuring the velocities of outer halo stars. 
\end{abstract}




\section{Introduction}\label{sec:introduction}


The Milky Way (MW) is undergoing a merger with the Large Magellanic Cloud \citep[LMC, see][for a comprehensive review of the effect of the LMC on the MW]{Vasiliev2023}. 
The LMC is thought to be on its first pericentric passage and to have a dark matter mass $M_{\mathrm{LMC}}\sim 10^{11}\,\mathrm{M}_{\odot}$ \citep{Besla2007, Besla2010, Boylan-Kolchin2011, Penarrubia2016, Kravtsov2024}.
An alternative scenario proposes the LMC is on its second pericentric passage; however, most observable features of this earlier passage scenario are superseded by the more recent, closer, pericentric passage \citep{Vasiliev2024}. 
Such a large mass for the LMC is required to explain a plethora of Local Group phenomena: for example, the kinematics of its globular clusters \citep{Watkins2024}, the kinematics of MW satellites \citep{Patel2020, CorreaMagnus2022, Kravtsov2024}; dynamical models of MW stellar streams \citep{2019MNRAS.487.2685E, Koposov2019, Shipp2021, Vasiliev2021, Koposov2023, Warren2025}; and the timing argument (\citealt{Penarrubia2016}, but see also \citealt{Benisty:2022, Chamberlain:2023, Benisty2024}) all require an LMC mass $M_{\mathrm{LMC}}\sim 1-2\, \times 10^{11}\,\mathrm{M}_{\odot}$ \citep[see fig.~1 of][for a summary of LMC mass estimates]{Vasiliev2023}. 
At present-day the LMC is at a distance of $d = (49.6 \pm 0.5) \,\mathrm{kpc}$ \citep{Pietrzynski2019},
and heliocentric line of sight velocity of $v_{\mathrm{los}} = (262.2 \pm 3.4)\,\mathrm{km}\,\mathrm{s}^{-1}$ \citep{vanderMarel2014}. 
An open question remains on the exact position, and hence velocity, of the LMC centre. A compilation of the recent reported LMC centre position and proper motion measurements is given in table~2 of \citet{Vasiliev2023}. 
The orbit of the LMC is sensitive to the assumed MW potential \citep[see fig.~3 of][]{Vasiliev2023} and,
because the LMC has a mass comparable to that of the MW \citep[$M_{\mathrm{MW}}\sim 10^{12}\,\mathrm{M}_{\odot}$][]{Wang2020}, it is also subject to dynamical friction from the MW dark matter halo \citep{Chandrasekhar1943}. 

The recent infall of the LMC into the MW generates a density wake in the MW dark matter halo \citep{Chandrasekhar1943}.
This occurs as the infalling LMC has a range of orbital frequencies which resonate with the orbits of dark matter particles of the MW's dark matter halo \citep{Mulder1983, Weinberg1986}. 
The classical `conic' wake trailing the LMC is typically labelled as the \textit{transient response}, whereas the response elsewhere in the MW halo is labelled the \textit{collective response} mainly caused by the shift in the systems' barycentre as well as containing some large scale resonance features \citep{Weinberg1989, Garavito-Camargo2019, Garavito-Camargo2021, Tamfal2021, Rozier2022, Foote2023}. 

The large mass ratio of the LMC to the MW, $\sim 10 - 20 \%$, combined with the fact that it has just completed its most recent pericentre passage at a relative velocity $ > 300 \,\mathrm{km}\,\mathrm{s}^{-1}$, has caused significant dynamical disequilibrium throughout the Galaxy \citep{Hunt2025}. 
In particular, the inner and outer parts of the MW halo have experienced different strengths of acceleration towards the LMC. To a Galactocentric observer, the Galactic northern sky appears to be red-shifted and the Galactic southern sky blue-shifted because the halo moves preferentially ‘up’, towards the Galactic north. 
This ‘reflex' displacement manifests as a dipole signal in density \citep{ Garavito-Camargo2021, Conroy2021, Amarante2024} that is higher in the Galactic North, and a dipole in stellar radial velocities \citep{2019MNRAS.487.2685E, 2020MNRAS.494L..11P, Petersen2021, Erkal2021, Yaaqib2024, Chandra2025a, Bystrom2025}.
Moreover, the density wake of the LMC is predicted to leave an observable signature in the density and kinematics of MW halo stars \citep[e.g.,][]{Belokurov2019, Conroy2021, Cavieres2025, Chandra2025a, Yaaqib2024, Bystrom2025, Fushimi24, Amarante2024, Sheng2025}.
The magnitude of the reflex velocity dipole is called the \textit{travel velocity} and its orientation is called the \textit{apex} direction of the reflex motion.
Recent studies have shown that the direction of the travel velocity vector points towards a point along the past orbit of the LMC; however, there is variance between studies on the exact preferred direction \citep[e.g.,][fig.~9]{Bystrom2025}. 
This is likely due to each study using different stellar tracers over varying radial ranges and sky coverages. 
A flexible and reliable technique which can account for differences between surveys will be powerful when performing inference for the reflex velocity, especially given the upcoming influx of outer halo datasets.

The MW--LMC system can be described by various levels of simulation fidelity. For any type of simulation, there exists a trade-off between the simulation fidelity and the ability to explore a large model parameter space. 
The simplest prescriptions are \textit{rigid} models of the MW and LMC galaxies. Rigid models describe the MW and LMC as analytic potentials that have fixed functional forms, i.e., they are time-invariant, although they are allowed to move in response to each other. 
The fidelity of the MW--LMC system can be increased by using $N$-body simulations in combination with Basis Function Expansions \citep[BFEs, e.g.,][often called deforming MW--LMC simulations]{Lilley2018a, Lilley2018b, Sanders2020, 2020MNRAS.494L..11P, Garavito-Camargo2019, Garavito-Camargo2021, Lilleengen2023, Vasiliev2024}. 
They aim to match the present-day conditions of the MW and LMC while also accounting for the dark matter halo responses of both galaxies during the infall of the LMC.
Furthermore, MW--LMC analogues have been identified in state-of-the-art cosmological hydrodynamical zoom-in simulations. For example, there are many examples within the \textit{Feedback In Realistic Environments} Latte \citep[\textit{FIRE},][]{Samuel2021, Wetzel2023, Garavito-Camargo2024, Arora2024}, the \textit{Milky Way-est} \citep{Buch2024}, the \textit{APOSTLE} \citep{Santos-Santos2021}, the \textit{Auriga} \citep{Grand2017, Smith-Orlik2023, Grand2024} and the \textit{DREAMS} \citep{Rose2025} simulation suites. Although cosmological analogues are inherently higher fidelity than $N$-body equivalents, they do not aim to exactly match present-day condition of the MW and LMC.

Simulation Based Inference \citep[SBI, see,][for a conceptual overview]{Cranmer2020} offers a medium to explore large and complex parameter spaces by producing many forward models of the system of interest.
SBI is a powerful statistical framework for performing inference in difficult modelling scenarios where traditional analytic methods are impractical or impossible. 
It is particularly well-suited for estimating parameter posterior distributions when the likelihood function cannot be explicitly defined and/or model parameter spaces are large and complex.
Instead of defining the likelihood, many forward simulations are used to generate samples of the data from a set of model parameters. 
SBI has been increasingly applied to a variety of astrophysics problems \citep[e.g.,][]{Weyant2013, Alsing2019, Jeffrey2021, Lemos2021, Hermans2021, vonWietersheim-Kramsta2024, Lovell2024, Ho2024, Widmark2025, Sante2025, XiangyuanMa2025, Jeffrey2025, Saoulis2025} and more generally across many fields of research e.g., seismology \citep[e.g.,][]{Saoulis2024}.
In \citet{Brooks2026}, we have previously shown for the MW--LMC system that inference on model parameters through an SBI framework trained on many, $\mathcal{O}(10^5)$, rigid simulations retains enough of the relevant physics that more complex simulations capture \citep[e.g., deforming $N$-body simulations,][]{Garavito-Camargo2019} to avoid model misspecification and vastly improve the computational efficiency of the inference. 

We present an SBI architecture for model parameter inference of the MW, LMC and the induced reflex motion using the measured velocities of outer MW halo stars. 
This builds upon the results of \citet{Brooks2026} where we demonstrated that an SBI framework trained on many rigid MW-LMC simulations can infer model parameters from dynamical measurements of outer halo tracers. 
The presented SBI architecture allows the exploration of large MW--LMC model parameter spaces, while incorporating time dependence, to enable the rapid and reliable inference of model parameters. 
For the first time, we demonstrate the application of our SBI architecture \citep{Brooks2026} to the Dark Energy Spectroscopic Instrument \citep[DESI,][]{DESICollaboration2025} Blue Horizontal Branch (BHB) and H3+SEGUE+MagE \citep{Chandra2025a} Red Giant Branch (RGB) datasets to provide constraints on the enclosed MW and LMC masses within $50\,\rm kpc$, the reflex motion and the strength of dynamical friction. 
Additionally, our framework allows us to explore the effects of survey sky coverage and availability of velocity information for stars in the outer halo.

The plan of the paper is as follows. 
In Sec.~\ref{sec:data} we describe the datasets used throughout this work and define the summary statistics used for model parameter inference.
In Sec.~\ref{sec:simulations}, we give a concise description of the MW--LMC simulations used to train the SBI architecture used for parameter inference.
In Sec.~\ref{sec:sbi}, we describe the SBI architecture, detailing the use of Bayesian statistics and the machine learning models used for parameter inference. 
In Sec.~\ref{sec:results}, we present and compare the constraints on the MW and LMC masses, the reflex motion and the dynamical friction model parameters.
In Sec.~\ref{sec:posterior-checks}, we perform a series of diagnostic tests on the estimated posterior distributions. 
In Sec.~\ref{sec:discussion}, we discuss our results and assess any model limitations. 
Finally, we conclude and provide an outlook for upcoming surveys in Sec.~\ref{sec:conclusions}.

\section{Data}\label{sec:data}

\begin{figure*}
    \centering
    \includegraphics[width=1\linewidth]{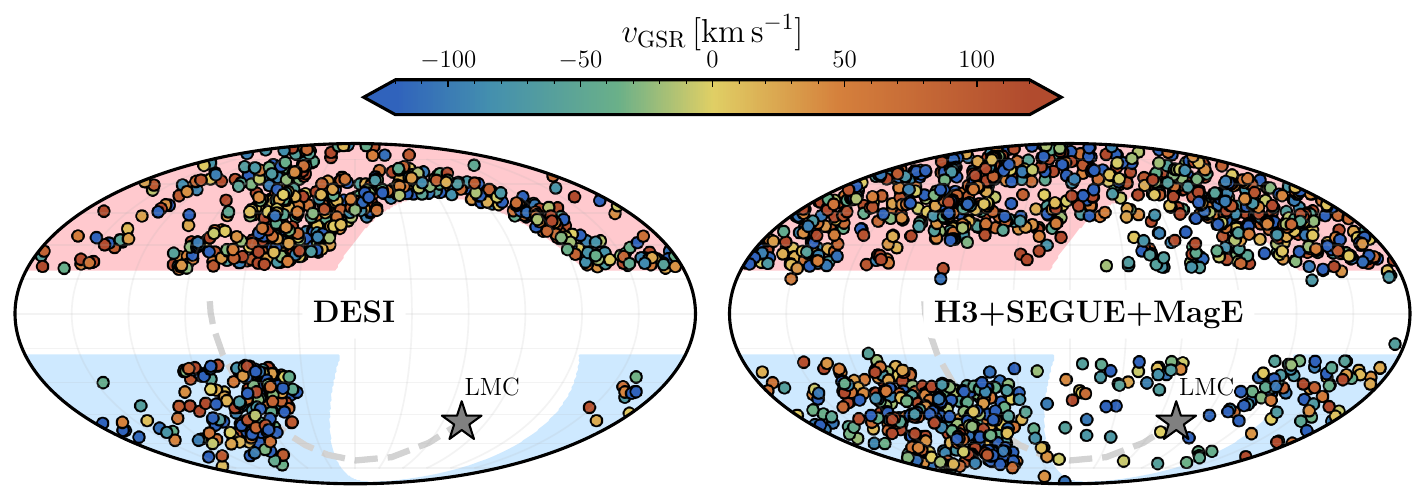}
    \includegraphics[width=.47\linewidth]{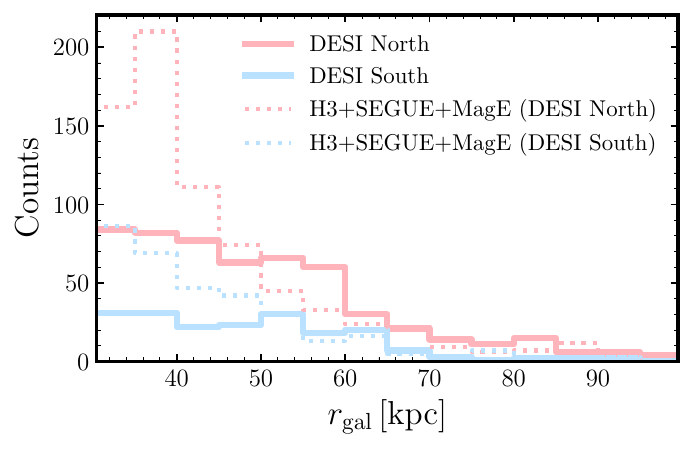}
    \caption{\textbf{Top panels:} The distribution of all sources beyond a Galactocentric distance of $30  \, \rm kpc$ for the Dark Energy Spectroscopic Survey (DESI, left) and the all-sky H3+SEGUE+MagE outer halo surveys (right) in Galactic coordinates. 
    The colour of individual sources reflects their solar corrected radial velocity, $v_{\rm GSR}$.
    The present-day position of the LMC is shown as the grey star along with an illustrative past orbit as the grey dashed line.
    The DESI northern (pink) and southern (blue) Galactic fields are shown as shaded regions.
    \textbf{Bottom panel:} The Galactocentric radial distribution of all sources between $30 - 100 \, \rm kpc$ for DESI (solid lines) and H3+SEGUE+MagE (dotted lines) contained within the DESI northern and southern Galactic fields.} 
    \label{fig:1}
\end{figure*}

To transform between the Heliocentric Cartesian and Galactocentric Cartesian coordinate frame in this work, we adopt a right-handed Cartesian coordinate system with the Sun positioned at $\vec{r}_{\odot} = (-8.3, 0.0, 0.02)\, \rm {kpc}$ \citep[][]{GRAVITYCollaboration2019, Bennett2019}, with
velocity $\vec{v}_{\odot} = (11.1, 244.24, 7.24)\, \rm {km}\,\rm s^{-1}$ \citep[][]{Schonrich2010, Eilers2019}. 

\subsection{Dark Energy Spectroscopic Survey}\label{sec:data-desi}

DESI is a multi-object spectrograph designed for ground-based wide-field surveys that operates on the Mayall 4-meter telescope at Kitt Peak National Observatory.
The DESI spectroscopic survey has a large sky coverage footprint of $14,000 \, \rm deg^{2}$. The instruments consist of $5,000$ fibres and cover a wavelength range of $360 - 960 \,\rm nm$ with a resolution between $2,000$ and $5,500$ depending on the wavelength \citep{DESICollaboration2022}. The main operating purpose of the survey is to obtain spectra for $\sim 40$ million galaxies and quasars to probe the nature of dark energy \citep{DESICollaboration2016}. Nevertheless, the DESI Milky Way Survey (MWS) working group also publishes the data of millions of individual stars in our Galaxy. In March 2025, the DESI Data Release 1 \citep[DR1,][]{DESICollaboration2025}\footnote{DESI Public Data Release 1: \url{https://data.desi.lbl.gov/doc/releases/dr1/}} was made available and included a catalogue containing over $6$ million MW sources with radial velocities and stellar parameters \citep{Koposov2025}. 
In this work, we use the Blue Horizontal Branch (BHB) star catalogue\footnote{The DESI BHB catalogue: \url{https://data.desi.lbl.gov/doc/releases/dr1/vac/mws-bhb/}} concurrently released by the MWS working group as a part of DESI DR1. 
The MWS provides radial velocities to all its targets via the radial velocity and stellar parameter fitting code \texttt{RVSpecfit} \citep{Koposov2011, Koposov2019}.

In DR1, there are a total of $10,695$ BHB targets, with the targeting procedure described in \citet{Cooper2023} and the precise distances to each BHB star is provided in \citet{Bystrom2025}.
We use the public BHB DESI DR1 catalogue, which has already been cleaned to ensure that known substructures, contaminating stars \& unphysical stellar quantities have been removed from the sample \citep[see,][sec.~2.4-2.6]{Bystrom2025}.
We use the \texttt{PRIMARY = True} flag to remove duplicate observations\footnote{Application of this flag exactly reproduces the BHB catalogue used in \citet{Bystrom2025}.}, and make a further selection to keep only stars between $r_{\rm gal} \in [30 - 100] \, \rm kpc$ to produce a final sample of $853$ DESI BHB stars for our analysis. 
In the upper left panel of Fig.~\ref{fig:1}, we show the on-sky distribution of these sources. Furthermore, in the lower panel of Fig.~\ref{fig:1} we show the number counts of sources as a function of Galactocentric distance, divided into the northern (pink) and southern (blue) DESI fields.
The observed on-sky and distance density distribution of DESI sources is non-uniform/linear. Although these effects may be small, from a simulation perspective, properly forward modelling MW stellar haloes should account for these effects. This is because SBI methods are sensitive to systematic biases if observational selection effects are not exactly modelled in the same way for the simulations used for inference. 

\subsection{H3, SEGUE \&  MagE Outer Halo Surveys}\label{sec:data-h3-segue-mage}

\begin{figure}
    \centering
    \includegraphics[width=0.9\linewidth]{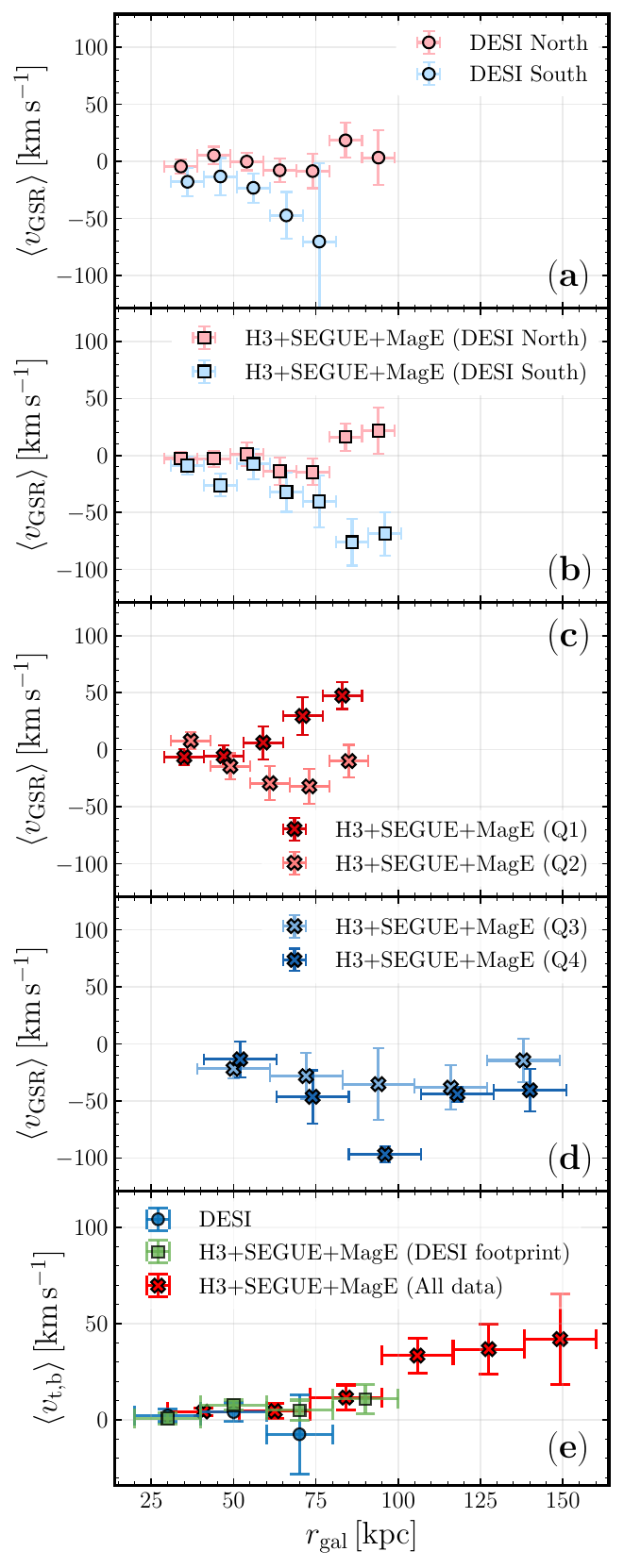}
    \caption{\textbf{Panel (a):} Mean radial velocity distributions, $\langle{v}_{\rm GSR} \rangle$, as a function of Galactocentric distance for the DESI sources in its northern (pink) and southern (blue) observing footprints. 
    \textbf{Panel (b):} Same as the first panel for H3+SEGUE+MagE sources within the DESI survey footprints.  
    \textbf{Panel (c/d):} The H3+SEGUE+MagE data divided into on-sky quadrant footprints.
    \textbf{Panel (e):} All-sky mean tangential velocity, $ \langle v_{\rm t, b} \rangle$,  distributions as a function of Galactocentric distance for DESI (blue), H3+SEGUE+MagE  within $100\,\rm kpc$ (green) and the full H3+SEGUE+MagE datasets (red). 
    Points are offset in distance for improved readability. All $1\sigma$ uncertainties are determined via bootstrap resampling.
    There is a clear increase in $\langle v_{\rm t, b} \rangle$ with distance out to $160\,\rm kpc$ for the H3+SEGUE+MagE data.} 
    \label{fig2}
\end{figure}

\subsubsection{H3}\label{sec:data-h3}

The H3 Spectroscopic Survey \citep{Conroy2019} has conducted a spectroscopic survey of halo stars
with the Hectochelle instrument \citep{Szentgyorgyi2011} on the $6.5\rm m$ MMT telescope at the Whipple Observatory in Arizona. 
We use the sample of H3 stars observed up to January 2024 that have reliable stellar parameters from \texttt{MINESweeper} \citep{Cargile2020} and are not associated with known MW substructures. 

\subsubsection{SEGUE}\label{sec:data-segue}

The Sloan Extension for Galactic Understanding and Exploration \citep[SEGUE,][]{Yanny2009} survey observed $\sim250,000$ stars with the low-resolution BOSS spectrograph as a part of the Sloan
Digital Sky Survey \citep[SDSS,][]{York2000}. 
These spectra have been fitted using the \texttt{MINESweeper} routine to provide reliable stellar parameters \citep{Chandra2025b, cargile_2025_16105186}.

\subsubsection{MagE}\label{sec:data-mage}

Over the past two years, a tailored spectroscopic survey of luminous RGB stars in the outer halo has been conducted with The Magellan Echellete Spectrograph \citep[MagE,][]{Marshall2008} on the $6.5\rm m$
Magellan Baade Telescope at Las Campanas Observatory.  
The selection procedure for the target sample of RGB stars is described in \citet{Chandra2023a}, and the details of the spectroscopic survey are further described in \citet{Chandra2023b, Chandra2025a}.
Stellar parameters are estimated with the \texttt{MINESweeper} code \citep{Cargile2020} including parallaxes measurements from Gaia \citep{GaiaCollaboration2023}.
As of May 2024, a total of $400$ stars have been observed, of which $\sim 300$ are spectroscopically confirmed to be at a heliocentric distance beyond $50 \, \rm kpc$, and $\sim 100$ are beyond $100 \, \rm kpc$ making it the largest dataset of outer halo stars beyond $50 \, \rm kpc$. For a much more detailed account of the MagE survey, we direct the reader to \citet{Chandra2023a, Chandra2023b, Chandra2025a}. 

\subsubsection{Combined H3+SEGUE+MagE sample}\label{sec:data-h3-segue-mage-sample}

The above surveys combine to produce a pure sample of stars with homogeneous stellar parameters derived with the \texttt{MINESweeper} pipeline.
Subsequently, we adopt the same selection procedure to exactly reproduce the high-fidelity subset in sec.~2.3, \citet{Chandra2025a}. This selection procedure ensures that known substructures \citep[e.g., the Sagittarius stream,][]{Majewski2003, Vasiliev2021} and unphysical quantities are removed from the sample.
The all-sky sample of H3+SEGUE+MagE stars used in this work contains $1296$ field stars between $r_{\rm gal} \in [30 - 100] \, \rm kpc$. 
In the upper right panel of Fig.~\ref{fig:1}, we show the on-sky distribution of these sources.
To ensure a fair comparison to the DESI survey, we apply a DESI sky coverage selection on this sample.
This DESI sky coverage sample contains $1049$ field stars.
We show their Galactocentric distance distribution in the lower panel of Fig.~\ref{fig:1}, divided into the northern (pink dotted) and southern (blue dotted) DESI fields.

\subsection{Observational summary statistics}\label{sec:data-measurements}

SBI requires a set of summary statistics to be used for the inference process. 
For this first application of an SBI framework to the dynamics of the outer MW stars, we adopt summary statistics based on their velocity field distributions. 
Specifically, we focus on the mean radial and tangential velocity distance distributions. 

\subsubsection{Radial velocities}\label{sec:data-measurements-vr}

We calculate the stellar radial velocities in the \textit{Galactic Standard of Rest} (GSR) frame, $v_{\rm GSR}$, which accounts for the solar motion with respect to the Galactic centre. 
For a non-rotating galaxy in equilibrium, the mean radial velocity, $\langle{v}_{\rm GSR} \rangle$, is expected to be zero throughout the entire MW. 
However, our Galaxy is in disequilibrium due to the merger with the LMC and hence $\langle{v}_{\rm GSR} \rangle \neq 0 \, \rm km\,\rm s^{-1}$ ubiquitously across the MW.

For both datasets, we calculate the $3 \sigma$-clipped mean $\langle{v}_{\rm GSR} \rangle$ values for the Galactocentric distance range $r_{\rm gal} \in [30 - 100] \, \rm kpc$ in bins of $10 \, \rm kpc$ width for the northern (pink) and southern (blue) DESI footprints; see panel (a, for DESI data) and panel (b, for H3+SEGUE+MagE data) in Fig.~\ref{fig2}. 
We choose this sky coverage and distance range so that the DESI and H3+SEGUE+MagE samples can be consistently compared, and to avoid any contamination in the inner halo, $r_{\rm gal} \lesssim 30 \, \rm kpc$. 
We note that both surveys reach more distant parts of the halo beyond $100\,\rm kpc$; however at these extreme distances the number of sources are limited. 
For this selection criteria, the remaining differences are intrinsic to the datasets themselves, e.g., measurement uncertainties and on-sky/distance density distributions. 

In addition to comparing DESI and H3+SEGUE+MagE in this way, we will also explore the full depth and on-sky coverage of the H3+SEGUE+MagE dataset by dividing the sky into quadrants and measuring a set of mean radial and tangential velocity summary statistics. 
Throughout this work, we define the quadrants as follows: Quadrant 1 (Q1) as $l \in [+180^{\circ}, 0^{\circ}]$, $b \in [0^{\circ}, +90^{\circ}]$, Quadrant 2 (Q2) as $l \in [0^{\circ}, -180^{\circ}]$, $b \in [0^{\circ}, +90^{\circ}]$, Quadrant 3 (Q3) as $l \in [+180^{\circ}, 0^{\circ}]$, $b \in [-90^{\circ}, 0^{\circ}]$ and Quadrant 4 (Q4) as $l \in [0^{\circ}, -180^{\circ}]$, $b \in [-90^{\circ}, 0^{\circ}]$; see the all-sky inset in Fig.~\ref{fig:app3} for a visual representation of these quadrants.
For the all-sky H3+SEGUE+MagE dataset, we calculate the $3 \sigma$-clipped mean $\langle{v}_{\rm GSR} \rangle$ values for the Galactocentric distance range $r_{\rm gal} \in [30 - 160] \, \rm kpc$ in 5 bins of equal width in the northern quadrants (Q1 and Q2) and the southern quadrants (Q3 and Q4); see panels (c/d) in Fig.~\ref{fig2}. 

Comparing the measured mean $\langle{v}_{\rm GSR} \rangle$ for the DESI and H3+SEGUE+MagE surveys using the consistent selection criteria (panels a/b in Fig.~\ref{fig2}), the general trends in the northern and southern DESI footprints are similar. 
The northern footprint displays behaviour almost consistent with expectations of dynamical equilibrium, whereas the southern footprint demonstrates an increasingly negative signal with increasing Galactocentric distance. 
For the DESI dataset, this is consistent with previous studies \citep[fig.~10,][]{Bystrom2025}.
For the H3+SEGUE+MagE data, we also see that the northern radial velocity distribution is consistent with expectations of dynamical equilibrium. 
This is a subtle difference to \citet{Chandra2025a} who showed an increasingly positive radial velocity amplitude in the northern Galactic hemisphere with Galactocentric distance.
This can be explained by the choice of sky coverage.
In \citet{Chandra2025a}, they selected the northern quadrant directly opposite the LMC (Q1) and a southern quadrant including the LMC (Q4) to measure the mean $\langle{v}_{\rm GSR} \rangle$ values. 
This choice of sky coverage, particularly in the northern Galactic hemisphere, is what drives the flattening of the mean $\langle{v}_{\rm GSR} \rangle$ distribution seen in the panels (c/d) of Fig.~\ref{fig2}. 
Although not the main result in this work, this insight nonetheless highlights the importance of survey sky coverage, particularly when using mean velocity values as summary statistics, as signal can be averaged out across larger areas. 

\subsubsection{Tangential velocities}\label{sec:data-measurements-vt}

The tangential velocity of a star, particularly in the Galactic latitude direction, $v_{t,b}$, traces the LMC’s perturbation in the MW \citep{Erkal2021, Sheng2024, Chandra2025a}. From the perspective of a Galactocentric observer, this component of the tangential velocity captures the apparent ‘upward’ reflex motion of outer halo stars as the MW's centre of mass is dragged ‘downward’.
For both datasets, we calculate the $3 \sigma$-clipped mean $\langle v_{t,b} \rangle$ values for the Galactocentric distance range $r_{\rm gal} \in [30 - 100] \, \rm kpc$ in distance bins of $20 \, \rm kpc$ width.
Plus, to exploit the full depth of the H3+SEGUE+MagE dataset we will also use the $3 \sigma$-clipped mean $\langle v_{t,b} \rangle$ values for the Galactocentric distance range $r_{\rm gal} \in [30 - 160] \, \rm kpc$ in 6 equally spaced distance bins.
In the panel (e) of Fig.~\ref{fig2} we show the all-sky mean tangential velocity distribution as a function of Galactocentric radius. 
These tangential velocities have been corrected for the solar motion. 
Both surveys demonstrate an increasing trend as a function of Galactocentric distance, albeit with DESI having very large uncertainties beyond $60\,\rm kpc$. Indeed, we do not show the outermost mean tangential velocity data point for DESI as the error bar is of the order $100\,\rm km \, s^{-1}$.

\section{Simulations}\label{sec:simulations}

In this section, we provide a concise description of the rigid MW--LMC simulations presented in \citet[][sec.~2]{Brooks2026}. 
We detail only the key information of these simulations and highlight any significant changes to the modelling. 
This rich set of low fidelity simulations, spanning a large model parameter space, is used to train the neural networks for parameter inference. 

\subsection{The Milky Way -- LMC potentials}\label{sec:simulations-MWLMCpot}

We use the galaxy dynamics C++/Python package \texttt{agama} \citep{2019MNRAS.482.1525V} to generate $128,000$ rigid MW--LMC simulations, each with a unique combination of model parameters.

\subsubsection{The Milky Way}\label{sec:simulations-MW}

To model the MW dark matter halo we use a Navarro-Frenk-White \citep[NFW,][]{Navarro1996, Navarro1997} dark matter halo density profile described by $M_{200}$ and $c_{200}$. 
These quantities are defined by a sphere enclosing an overdensity that is 200 times the critical density of the Universe, $\rho_{\mathrm{crit}} = 3H_{0}^{2}/8\pi G$, as denoted by the ‘$200$' subscript, where Hubble parameter, $H_0$, is taken to be $67.6\, \rm km\,\rm s^{-1} \, \rm Mpc^{-1}$ using the default cosmology in \texttt{astropy} \citep{astropy:2013}.
We constrain the normalisation of the halo mass profile such that the circular velocity at the solar position is approximately $235\,\mathrm{km}\,\mathrm{s}^{-1}$ \citep[e.g., matching constraints from][within the associated uncertainty]{McMillan2017}.
The MW stellar components are modelled using a spherical bulge with a total mass $1.2 \times 10^{10} \, \mathrm{M}_{\odot}$, and an exponential stellar disc with a total mass $5 \times 10^{10} \, \mathrm{M}_{\odot}$. 
The stellar distributions remain fixed taking the values suggested by \cite{McMillan2017}. 

\subsubsection{The LMC}\label{sec:simulations-lmc}

We model the LMC as a Hernquist dark matter halo \citep{Hernquist1990}. 
We normalise the profile such that the derived rotation curve peaks at $(91.7 \pm \, 18.8) \,\mathrm{km}\,\mathrm{s}^{-1}$ at a distance of $8.7\,\mathrm{kpc}$ from its centre. This corresponds to an enclosed dynamical mass of $M_{\rm LMC}(r<8.7\,\mathrm{kpc}) = (1.7 \pm 0.7) \times 10^{10}\,\mathrm{M_{\odot}}$ \citep{vanderMarel2014} . 

\subsubsection{The MW--LMC interaction}\label{sec:simulations-mwlmcinteraction}

The trajectories of the MW and LMC under their mutual gravitational attraction are numerically integrated \citep[see equ.~3-6 in][]{Brooks2026}. 
We account for acceleration due to Chandrasekhar dynamical friction, $\mathbf{a}_{\mathrm{DF}}$, on the trajectory of the LMC \citep{Chandrasekhar1943, Binney2008, Jethwa2016} as:

\begin{equation}\label{equ1}
\mathbf{a_{\rm DF}} = - \frac{4 \pi G^2 M_{\rm LMC} \rho_{\rm MW} \ln \Lambda}{v_{\rm LMC}^3} 
\left[ \operatorname{erf}(X) - \frac{2X}{\sqrt{\pi}} e^{-X^2} \right] \mathbf{v_{\rm LMC}} \times \lambda_{\mathrm{DF}},
\end{equation}
\noindent
where $X = v_{\rm LMC} / \sqrt{2}\sigma_{\rm MW}$ and $\sigma_{\rm MW}$ and $\rho_{\rm MW}$ are the velocity dispersion and total density field of the MW. 
Following \citet{Vasiliev2021}, we take a fixed value of $\sigma_{\rm MW} = 120 \,\rm km \, \rm s^{-1}$ for the velocity dispersion as the dynamical friction is insensitive to the precise value. 
For the Coulomb logarithm we adopt \(\ln \Lambda = \ln \left( 100\, \rm kpc / \epsilon \right)\). The softening length, $\epsilon$, depends on the satellite’s density profile \citep{White1976}. 
We adopt $\epsilon = 1.6 \, a_{\rm LMC}$ as this has been used previously when modelling the LMC as a Plummer sphere \citep[e.g.,][]{Hashimoto2003, Besla2007, vanderMarel2012, Sohn2013, Kallivayalil2013}. 
The numerator in the Coulomb logarithm expression an arbitrarily chosen value that loosely describes the average separation of the MW and LMC. 
In principle this value could be updated through the integration of the LMC orbit, however this will have a small effect. 
We use a dimensionless parameter, $\lambda_{\mathrm{DF}}$, to modulate the strength of the dynamical friction that the LMC experiences. In principle, this will take into account changes to the fixed Coulomb logarithm value per simulation. 
The final MW--LMC potential includes the acceleration of the MW's centre of mass towards the LMC. 

\subsection{The Milky Way Stellar Halo}\label{sec:simulations-stellarhaloes}

To generate a mock MW stellar halo for each simulation, we draw phase-space samples from radially-biased distribution functions as implemented in \texttt{agama} \citep{2019MNRAS.482.1525V}.
This requires instances of a tracer density profile, a potential, and a prescription for the radial velocity anisotropy. 
We use a Dehnen tracer density profile \citep{Dehnen1993} and an NFW profile for the potential \citep{Navarro1996, Navarro1997} with relevant parameters adopted from each unique MW--LMC simulation.  
We implement the radial bias of stellar velocities \citep{Osipkov1979, Merritt1985, Binney2008} using the velocity anisotropy profile:
\begin{equation}
    \beta(r) \equiv 1 - \frac{\sigma_t^2}{2\sigma_r^2} = \frac{\beta_0 + (r/r_a)^2}{1 + (r/r_a)^2}.
\end{equation}
\noindent
where $\beta_0$ is the limiting value of anisotropy in the centre, and if $r_a < \infty$, the anisotropy coefficient tends to 1 at large $r$, otherwise it is constant and equal to $\beta_0$ over all scales which we adopt. 

A key improvement from our previous stellar halo simulations in \citet{Brooks2026} is that we sample $\sim 4.5$ times more phase-space coordinates to initialise the MW stellar haloes before any LMC-induced disequilibrium.
This $\sim 4.5$-fold increase in the number of particles, $20,000$ for each stellar halo, used to represent the stellar halo ensures that Poisson noise does not dominate the uncertainty on measurable quantities. 
Thus the main source of uncertainty is from the observations themselves.
With this increased measurement precision from our simulations, we can convolve any measured value with the observational error from any given survey. 
This allows us to correctly forward model the simulations to produce observational-like quantities which are subsequently used for the evaluation of the posterior.
Hence, prior to training the inference framework, we apply the survey-specific uncertainties to the binned radial and tangential velocity measurements from the DESI and H3+SEGUE+MagE datasets; see Sec.~\ref{sec:data-measurements}. 
This approach more closely aligned with SBI ideologies, allowing us to better forward model all the stellar haloes to match a specific survey of interest and perform the subsequent inference.
For a given MW--LMC potential with reflex motion, we integrate all particles in the stellar halo to present-day over the last $2.2\,\mathrm{Gyr}$. 

From the final distribution of stellar halo particles, we measure the reflex motion of the MW in response to the infalling LMC \citep[as used in][]{Petersen2021, Chandra2025a, Yaaqib2024, Yaaqib2025, Brooks2026}.
To do this, we use all stars beyond $50 \, \rm kpc$.
Although, it is worth noting that this calculation from the dynamics of outer halo stars is not necessarily exactly the same as the induced velocity of the MW centre relative to the initial inertial frame prior to the LMC's infall.
This method fits an on-sky velocity model which contains nine free parameters. We model the dipole reflex motion using Galactocentric Cartesian velocities $\{v_x, v_y, v_z \}$. 
We account for non-zero mean motion in the halo's Galactocentric velocity via the mean motion parameters $\vec{v}_{\mathrm{mean}} = \left( \langle v_r \rangle, \langle v_{\phi} \rangle, \langle v_{\theta} \rangle \right)$. This allows for any departures in the bulk halo motion from the travel velocity.
Finally, we account for the intrinsic velocity dispersion in each component using the set of hyperparameters, $\{ \sigma_{v_{\rm r}},\sigma_{v_l}, \sigma_{v_b} \}$, which account for measurement uncertainties. 
The reflex motion model is represented by the sum of the dipole and mean motion parameters:
\begin{equation}
    \langle \vec{v}\rangle = \vec{v}_{\rm travel} + \vec{v}_{\mathrm{mean}}, 
\end{equation}
\noindent
where $\vec{v}$ is the mean Galactocentric halo velocity vector. 
To find the maximum likelihood estimates for these parameters given each set of mock stellar halo data, we minimise a Gaussian log-likelihood for the 1-dimensional line of sight velocities and 2-dimensional proper motions using \texttt{scipy.optimize} \citep[see equ.~6 \& 8 in][]{Petersen2021}. 
We return the maximum likelihood estimates for all of the reflex motion model parameters. 
However, in the context of this work, we will only comment on the magnitude, $v_{\rm travel}$, and Galactocentric components of the reflex travel velocity, $\{v_x, v_y, v_z \}$.

\subsection{Simulation Priors}\label{sec:simulations-ics}

We use the same parameter prior probability distributions as in \citet{Brooks2026}. For clarity, we repeat that information here in Table.~\ref{table:priors}.
The first two parameters are the MW total mass and enclosed mass within $50\,\rm kpc$.
The next two parameters are the infall LMC total mass and its enclosed mass within $50\,\rm kpc$.
Next is the scalar multiple that modulates the strength of dynamical friction relative to classic Chandrasekhar values; see Equ.~\ref{equ1}.
The next set of parameters describe the LMC present-day position and velocity with their distributions inspired by the values in sec.~3.1 and table~2 of \citet{Vasiliev2023}. 
The final two parameters are the anisotropy parameter, $\beta_0$, and the Dehnen tracer density profile scale length, $r_{\rm Dehnen}$, that initialise the mock MW stellar haloes.
In total, we run $128,000$ MW--LMC simulations each with unique parameter values and $20,000$ particles to represent the MW stellar halo.

\begin{table}
\centering
\caption{Simulation model parameter prior distributions. The mass enclosed priors (grey) are derived using the priors for the total masses.}
\label{table:priors}
\begin{tabularx}{\linewidth}{lX}
\hline
\hline
Model Parameter & Prior probability distribution \\
\hline
& \\
\( M _{200,\rm MW} \) & \(  \mathcal{N}(15, 5) \times 10^{11} M_{\odot} \) \\
\textcolor{gray}{\( M _{\rm MW}(< 50\,\mathrm{kpc}) \)} & 
\textcolor{gray}{\(  \mathcal{N}(4.8, 0.8) \times 10^{11} M_{\odot} \)} \\
& \\
\(  M_{\rm LMC} \) & \( \mathcal{N}(15, 10) \times 10^{10} M_{\odot} \) \\
\textcolor{gray}{\( M _{\rm LMC}(< 50\,\mathrm{kpc}) \)} & \textcolor{gray}{\(\mathcal{N}(8.6, 2.9) \times 10^{10} M_{\odot} \)} \\
& \\
\( \log_{10}(\lambda_{\mathrm{DF}}) \) & \( \mathcal{U}(-3, 1) \) \\
& \\
\( \alpha_{\rm LMC} \) & \(\mathcal{U} (60^\circ, 90^\circ) \) \\
\( \delta_{\rm LMC} \) & \(\mathcal{U}(-80^\circ, -50^\circ)  \) \\ 
\( d_{\rm LMC} \) & \( \mathcal{N}(49.6, 5)\,\mathrm{kpc}\) \\
\( v_{\rm los} \) & \( \mathcal{N}(262.2, 10) \,\mathrm{km}\,\mathrm{s}^{-1} \) \\
\( \mu_{\alpha_{\rm LMC}} \) & \( \mathcal{N}(1.9, 0.25)\, \mathrm{mas}\,\mathrm{yr}^{-1} \) \\ 
\( \mu_{\delta_{\rm LMC}} \) & \( \mathcal{N}(0.33, 0.25)\,\mathrm{mas}\,\mathrm{yr}^{-1} \) \\
& \\
\(\beta_0 \) & \(  \mathcal{U}(0, 0.9)\) \\
\( r_{\rm Dehnen} \) & \(\mathcal{U}(10, 15)\, \rm kpc \) \\
\hline
\end{tabularx}
\end{table}

\section{Simulation Based Inference}\label{sec:sbi}

In the Bayesian approach, a problem is often posed as calculating the probability of the model
parameters $\theta$, given some observed data $D_{\text{obs}}$, and a theoretical model $I$. In other words, we want to find the posterior probability distribution, $\mathcal{P} = p(\theta | D_{\text{obs}}, I)$. This is possible using Bayes' Theorem:

\begin{equation}\label{equ4}
     p(\theta | D_{\text{obs}}, I) = \frac{p(D_{\text{obs}} | \theta, I) p(\theta | I)}{p(D_{\text{obs}} | I)}
\iff \mathcal{P} = \frac{\mathcal{L} \times \Pi}{\mathcal{Z}} 
\end{equation}
\noindent
where $\mathcal{L} = p(D_{\text{obs}} | \theta, I)$ is the likelihood, $\Pi =   p(\theta | I)$ is the prior, and $\mathcal{Z} = p(D_{\text{obs}} | I)$ is the
Bayesian evidence. The Bayesian evidence acts as a normalisation in parameter estimation and can be ignored for our application. Given a choice of prior distribution for parameters and a likelihood function, we can find the posterior distribution. In the case where a likelihood
function need not, or cannot be explicitly defined, it is possible to instead use SBI to estimate the posterior distribution. 

The simplest form of SBI is known as Approximate Bayesian Computation \citep[ABC, e.g.,][]{Rubin1984, Pritchard1999, Fearnhead2010}. The ABC framework selects forward simulations that are the most similar to the observed data based on some distance measure involving the summary statistics of the simulation.
Another way to compute the posterior is via Density Estimation Likelihood Free Inference (DELFI). In this approach, forward simulations are used to learn a conditional density distribution of the data $D_{\text{obs}}$, given the simulation parameters $\theta$, using a density estimation algorithm, e.g., normalising flows, that utilise a series of bijective transformations to convert a simple base distribution e.g., Gaussian, into the target probability distribution \citep{JimenezRezende2015}. 
We adopt a DELFI approach using the \texttt{sbi} Python package \citep{tejero-cantero2020sbi}, and estimate the posterior distribution from the forward simulations using Masked Autoregressive Flows \cite[MAF,][]{Papamakarios2017, Papamakarios2019} with $5$ transformation layers in the neural network, each with a width of $50$ nodes. 
The exact neural network architecture used can influence the estimated posterior. 
However, we have found that varying the number of layers and nodes has very little effect on our results. 
DELFI is advantageous over the simpler ABC approach as it does not rely on a choice of a distance measure and it uses all available forward simulations to build the posterior distribution, making it far more efficient \citep{Alsing2019}. 
Once a normalising flow has been trained on a precomputed simulation dataset, the posterior can be returned for many observations without having to retrain the flow; this is known as \textit{amortisation} \citep{Mittal2025}.

Often, some form of data compression is required \citep[e.g.,][]{Alsing2018, Alsing2019b, Heavens2020, Jeffrey2021, Widmark2025, Jeffrey2025}. However, the application to our problem is relatively low in dimensionality, i.e., the number of parameters of interest and data points, hence no data compression is required. 
Generally, for SBI, the more simulations that are available to use, the better. Within a cosmological context, the estimated number of simulations that are required for reliable SBI analysis is $\sim 10^{4}$ \citep{Bairagi2025}. 

The simulations used in this work are described in Sec.~\ref{sec:simulations}. 
We use a MAF density estimator from the \texttt{sbi} package \citep{tejero-cantero2020sbi} to directly obtain the posterior distribution that can be evaluated at any observed data point for any data realisation, i.e, \( p(\theta | D_{\text{obs}}, I) \).
We ensure the reliability of the estimated posteriors through some diagnostic checks including coverage probabilities and predictive posterior checks in Sec.~\ref{sec:posterior-checks}.

\section{Results}\label{sec:results}

We show posterior distributions for MW, LMC and reflex motion model parameters conditioned on different data subsets.
All figures use $10,000$ samples drawn from their respective posterior density distributions.

\subsection{Inference set-up}

Throughout this section, we return the posterior distributions for MW and LMC masses enclosed within $50\,\rm kpc$, the Cartesian components of the travel velocity and the dynamical friction strength. 
Following the Bayesian notation in Equ.~\ref{equ4}, this reads as: \begin{equation*}
    \theta = \{ M_{\mathrm{MW}}(< 50 \, \mathrm{kpc}),  M_{\mathrm{LMC}}(< 50 \, \mathrm{kpc}),  v_x, v_y, v_z, \log \lambda_{\rm{DF}}\} 
\end{equation*}
\noindent
For the input stellar velocity data used to condition the posterior on, we will make it clear in each sub-section which on-sky selection and dataset is used. 
In Sec.~\ref{sec:results-desifoot}, for sources that lie within the DESI survey footprint boundaries, we present the constraints obtained when using either the DESI or H3+SEGUE+MagE datasets.
In Sec.~\ref{sec:results-quadfoot}, we use an on-sky selection defined by four quadrant footprints, and explore the full depth of the H3+SEGUE+MagE dataset.
We note that this exploration of using quadrants is only possible for all-sky datasets like H3+SEGUE+MagE. 
Therefore the same analysis cannot be carried out using a limited sky coverage survey like DESI.
As a reminder, observable uncertainties are accounted for in our simulated stellar halo velocity measurements; see Sec.~\ref{sec:simulations-stellarhaloes} for details on accounting for survey specific uncertainties.
Finally, the only observed data points that are consistently used for each posterior estimation are the present-day Galactocentric position and velocity of the LMC as $\vec{x}_{\rm LMC} = 
\{-0.6, -41.3, -27.1 \} \, \rm kpc$, $\vec{v}_{\rm LMC} = \{-63.9, -213.8, 206.6 \} \, \rm km \, s^{-1}$ \citep[][and references therein]{Vasiliev2021}.
Uncertainty in the LMC position and velocity is accounted for by convolving each simulated value with the uncertainties given in sec.~3.1 and table~2 of \citet{Vasiliev2023}. 
In our models, the true present-day LMC phase space coordinates are treated as model parameters, with the observed phase space coordinates, i.e. $(\vec{x}_{\rm LMC}, \vec{v}_{\rm LMC})$ from \citet{Vasiliev2021}, providing noisy measurements. One could infer a posterior over these parameters, though they are well constrained and not the focus of this work, so we choose to condition our inference on their observed values within the explored prior space.

\subsection{DESI footprint}\label{sec:results-desifoot}

This section only uses the data from sources that are contained within the DESI northern and southern survey footprints. 
This is described in detail in Sec.~\ref{sec:data-measurements}, and the measurements for the binned velocity field summary statistic that are used as input data are shown in Fig.~\ref{fig2}. 
Following the Bayesian notation in Equ.~\ref{equ4}, the input data reads as: 
\begin{gather*}
        D_{\text{obs}} = \{ (\vec{x},\vec{v})_{\rm LMC}, \langle v_{\mathrm{GSR, N}} \rangle, \langle v_{\mathrm{GSR, S}} \rangle \}, \,\rm or  \\
        D_{\text{obs}} = \{ (\vec{x},\vec{v})_{\rm LMC}, \langle v_{\mathrm{GSR, N}} \rangle, \langle v_{\mathrm{GSR, S}} \rangle, \langle v_{\rm t, b} \rangle \},  
\end{gather*}
\noindent
as we vary whether or not the tangential velocity data points, $\langle v_{\rm t, b} \rangle$, are provided. The subscripts N and S signify the binned radial velocity data, $\langle v_{\mathrm{GSR}} \rangle$, in the northern and southern DESI survey fields, respectively. 

\subsubsection{DESI data}\label{sec:results-desifoot-desidata}

\begin{figure}
    \centering
    \includegraphics[width=\linewidth]{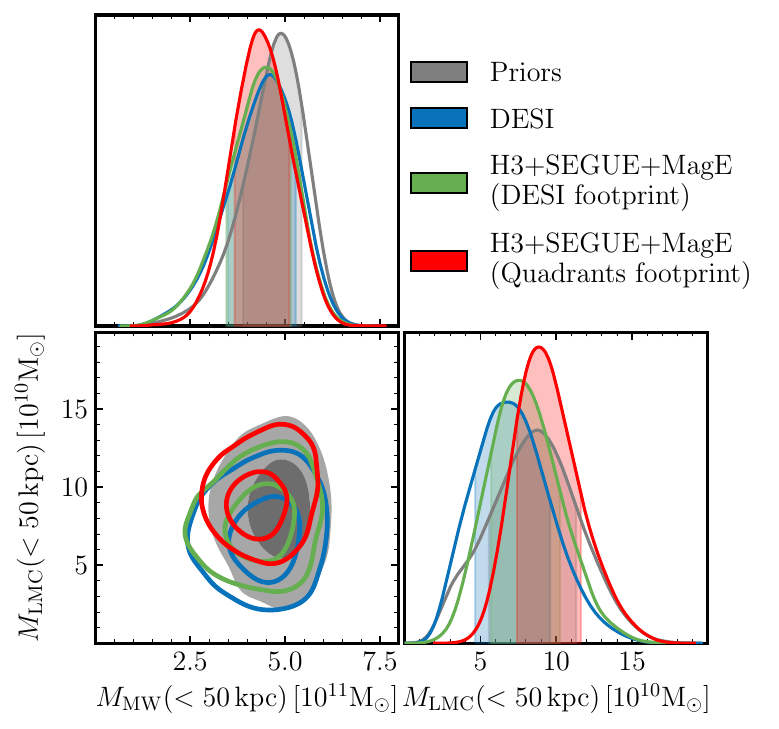}
    \caption{\textbf{Posterior distributions - MW and LMC enclosed masses:} The joint, and individual, posterior distributions for the MW and LMC masses enclosed within $50\,\rm kpc$. 
    We show these distributions conditioned on the LMC centre present-day position \& velocity, and radial and tangential velocities as data points.
    The open blue, green and red contours represent the posteriors conditioned using the DESI, H3+SEGUE+MagE (DESI footprint) and H3+SEGUE+MagE (Quadrants footprint) survey data, respectively.
    The prior distributions are shown as the filled grey contours.
    The contours delineate the $1\sigma$ and $2\sigma$ confidence intervals. 
    For the 1D posterior panels we show the $16^{\rm th}-84^{\rm th}$ percentiles as shaded regions.}
    \label{fig3}
\end{figure}

The measured data for the binned radial velocities is shown in panel (a) of Fig.~\ref{fig2} and the binned tangential velocities are shown as the blue points in the panel (e) of Fig.~\ref{fig2}.
We show the posteriors when providing information on the LMC present-day position, velocity and the radial and tangential velocities as the open blue contours in Fig.~\ref{fig3}, for the MW and LMC enclosed masses, and in Fig.~\ref{fig4} for the Galactocentric cartesian components of the travel velocity. 
In Appendix~\ref{sec:appendixA}, we show the full posterior distribution for the model parameter constraints using the DESI dataset in Fig.~\ref{fig:app1}.
This includes the results when we only provide the radial velocities information for the outer halo tracers.
We show the model priors, or simulation values, as the filled grey contours.
The joint posterior distributions show the $1\sigma$ and $2\sigma$ confidence intervals. 
The individual posterior distributions show the $16^{\rm th}$ and $84^{\rm th}$ percentiles as filled bands.
For all parameters shown, the inclusion of DESI tangential velocity information does not greatly improve the precision of the inferred parameters. 
This is not unexpected as the uncertainties on the measured values are very large; see panel (e) of Fig.~\ref{fig2}.
We summarise these results in the top two rows of Table.~\ref{table:posterior-results} and compare them to previous results in Sec.~\ref{sec:results-comparison}.

\begin{figure*}
    \centering
    \includegraphics[width=\linewidth]{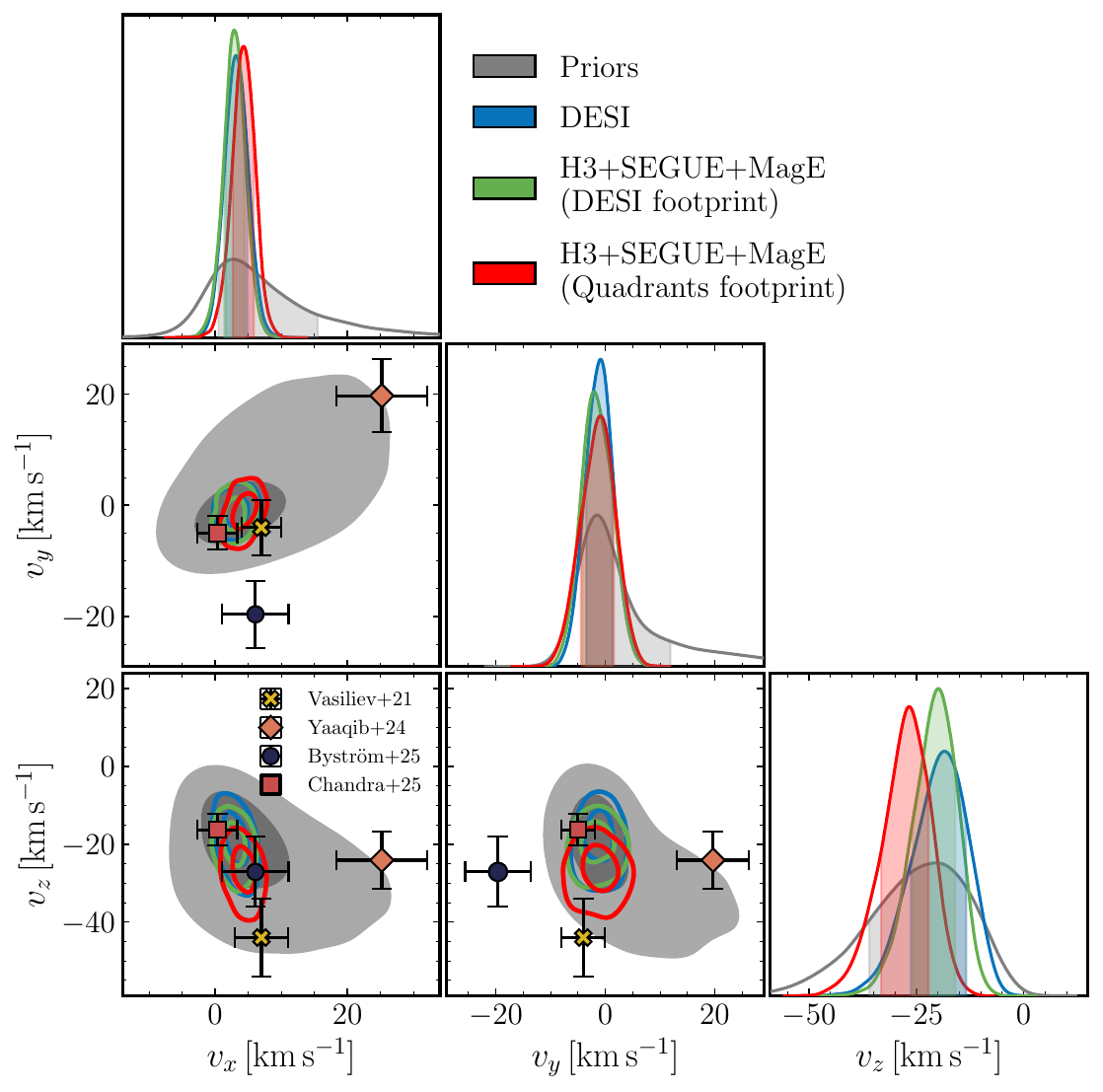}
    \caption{\textbf{Posterior distributions - Reflex motion velocity:} The joint, and individual, posterior distributions of the Galactocentric Cartesian travel velocity components.
    We show these distributions conditioned on the LMC centre present-day position \& velocity, and radial and tangential velocities as data points.
    The open blue, green and red contours represent the posteriors conditioned using the DESI, H3+SEGUE+MagE (DESI footprint) and H3+SEGUE+MagE (Quadrants footprint) survey data, respectively.
    The contours delineate the $1\sigma$ and $2\sigma$ confidence intervals.
    For the 1D posterior panels we show the $16^{\rm th}-84^{\rm th}$ percentiles as shaded regions.
    The measured mean and $1\sigma$ errors from \citet[][yellow cross]{Vasiliev2021}, \citet[][orange diamond]{Yaaqib2024}, \citet[][blue circle]{Bystrom2025} and \citet[][red square]{Chandra2025a} are shown in each panel for comparison.}
    \label{fig4}
\end{figure*}

\subsubsection{H3+SEGUE+MagE data}\label{sec:results-desifoot-h3data}

The measured data for the binned radial velocities is shown in panel (b) of Fig.~\ref{fig2} and the binned tangential velocities are shown as the green squares in panel (e) of Fig.~\ref{fig2}.
We show the posteriors when providing information on the LMC present-day position, velocity and the radial and tangential velocities as the open green contours in Fig.~\ref{fig3}, for the MW and LMC enclosed masses, and in Fig.~\ref{fig4} for the Galactocentric cartesian components of the travel velocity. 
In Appendix~\ref{sec:appendixA}, we show the full posterior distribution for the model parameter constraints using the H3+SEGUE+MagE dataset in Fig.~\ref{fig:app2}.
This includes the results when we only provide the radial velocities information for the outer halo tracers.
We summarise these constraints in the middle two rows of Table.~\ref{table:posterior-results}.
The constraints on all model parameters returned using DESI or H3+SEGUE+MagE data contained within the DESI survey footprints have similar values and are consistent within uncertainties. We get slightly more precise constraints on the travel velocity and enclosed LMC mass, $\sim 10\%$, using the H3+SEGUE+MagE dataset, although this is expected as there are more sources compared to DESI.
We make further comparisons to previous results in Sec.~\ref{sec:results-comparison}.

\subsection{Quadrants footprint}\label{sec:results-quadfoot}

In this section, we explore the full H3+SEGUE+MagE dataset to extract the maximum constraining power by dividing the sky into quadrants and measuring a set of mean radial and tangential velocity summary statistics. 
A visual representation of these quadrants is shown in the all-sky inset of Fig.~\ref{fig:app3}.
We use all H3+SEGUE+MagE sources between $30$ and $160\,\rm kpc$.
The measured data for the binned radial velocities is shown in panels (c/d) of Fig.~\ref{fig2} and the binned tangential velocities are shown as the red points in the panel (e) of Fig.~\ref{fig2}.
The distance bins used to measure the average radial and tangential velocity fields are the same as in \citet[][figures 4 \& 5]{Chandra2025a}.
Utilising an on-sky quadrant footprint is only possible for all-sky datasets. Hence we cannot carry out a similar analysis using the DESI data. 
In principle, as the H3+SEGUE+MagE dataset contains a larger sample of stars that have been observed deeper in the MW halo, and allows for the finer on-sky division, we expect these results to be the most precise of all those we present in this work.

Following the Bayesian notation in Equ.~\ref{equ4}, the input data reads as: 
\begin{gather*}
        D_{\text{obs}} = \{(\vec{x},\vec{v})_{\rm LMC}, \langle v_{\mathrm{GSR, Q1}} \rangle,\, \langle v_{\mathrm{GSR, Q2}} \rangle, \langle v_{\mathrm{GSR, Q3}} \rangle,\, \langle v_{\mathrm{GSR, Q4}} \rangle\} \\ \mathrm{or,}  \\
        D_{\text{obs}} = \{(\vec{x},\vec{v})_{\rm LMC}, \langle v_{\mathrm{GSR, Q1}} \rangle,\, \langle v_{\mathrm{GSR, Q2}} \rangle, \langle v_{\mathrm{GSR, Q3}} \rangle,\, \langle v_{\mathrm{GSR, Q4}} \rangle, \\
        \hspace{1.1cm} \langle v_{\rm t, b} \rangle \},  
\end{gather*}
\noindent
as we vary whether or not the tangential velocity data points, $\langle v_{\rm t, b} \rangle$, are provided. The numerical subscripts signify the binned radial velocity data, $\langle v_{\mathrm{GSR}} \rangle$, in each respective on-sky quadrant.

Similar to before, we show the posteriors when providing information on the LMC present-day position, velocity and the radial and tangential velocities as the open red contours in Fig.~\ref{fig3}, for the MW and LMC enclosed masses, and in Fig.~\ref{fig4} for the Galactocentric cartesian components of the travel velocity. 
In Appendix~\ref{sec:appendixA}, we show the full posterior distribution for the model parameter constraints using the H3+SEGUE+MagE dataset in Fig.~\ref{fig:app2}.
This includes the results when only provide the radial velocities information for the outer halo tracers.
We show the prior distributions as the filled grey contours.
The joint posterior distributions show the $1\sigma$ and $2\sigma$ confidence intervals. 
The individual posterior distributions show the $16^{\rm th}$ and $84^{\rm th}$ percentiles as filled bands.
We see improvement in the precision of the inferred parameters, $\sim 15-20\%$, by including the tangential velocity information. 
Although, the parameters are already well constrained using the radial velocity information alone.
We summarise these results in the final two rows of Table.~\ref{table:posterior-results} and compare them to previous results in Sec.~\ref{sec:results-comparison}.

\setlength{\tabcolsep}{7pt}
\renewcommand{\arraystretch}{1.3} 
\begin{table*}
\centering
\caption{The posterior medians and $1\sigma$ credible intervals for the magnitude of the reflex motion travel velocity, $v_{\rm travel}$, its Galactocentric Cartesian vector components $v_x$, $v_y$, $v_z$ (units, $\rm km \, s^{-1}$), the MW (units, $ 10^{11} \, \rm M_{\odot}$) and LMC (units, $ 10^{10} \, \rm M_{\odot}$) masses enclosed within $50\,\rm kpc$ and the strength of Chandrasekhar dynamical friction, $\log \lambda_{\rm DF}$ (units, dimensionless). The dataset and footprint used to produce each constraint are shown in the leftmost column.}
\begin{tabular}{lccccccc}
\hline
\hline
Data \& Footprint & $v_{\rm travel} $ & $v_x $ & $v_y $ & $v_z $ & $M_{\rm LMC}(< 50 \, \rm kpc) $ & $M_{\rm MW}(< 50 \, \rm kpc)$ & $\log \lambda_{\rm DF}$ \\
\hline
\vspace{-0.3cm} \\
\shortstack[l]{\textbf{DESI data}, \\ DESI footprint, without $v_{t,b}$} & $20.4^{+7.5}_{-6.2}$ & $3.2^{+2.0}_{-1.9}$ & $-1.0^{+2.7}_{-2.7}$ & $-20.0^{+6.3}_{-7.6}$ & $7.3^{+2.8}_{-2.5}$ & $4.4^{+0.8}_{-1.0}$ & $0.1^{+0.6}_{-1.6}$ \\
\shortstack[l]{\textbf{DESI data}, \\ DESI footprint, with $v_{t,b}$} & $19.7^{+6.8}_{-5.7}$ & $3.2^{+1.7}_{-1.7}$ & $-1.1^{+2.5}_{-2.6}$ & $-19.2^{+5.8}_{-6.8}$ & $7.0^{+2.7}_{-2.4}$ & $4.5^{+0.8}_{-1.0}$ & $0.2^{+0.6}_{-1.7}$\\
\shortstack[l]{\textbf{H3+ data}, \\ DESI footprint, without $v_{t,b}$} & $24.9^{+6.7}_{-5.7}$ & $3.8^{+1.8}_{-1.7}$ & $-0.5^{+3.1}_{-3.1}$ & $-24.4^{+5.8}_{-6.6}$ & $8.9^{+2.5}_{-2.2}$ & $4.5^{+0.8}_{-0.7}$ & $0.1^{+0.7}_{-1.7}$\\
\shortstack[l]{\textbf{H3+ data}, \\ DESI footprint, with $v_{t,b}$} & $20.9^{+5.6}_{-4.7}$ & $2.9^{+1.6}_{-1.6}$ & $-1.5^{+2.9}_{-2.7}$ & $-20.4^{+4.7}_{-5.6}$ & $7.8^{+2.5}_{-2.2}$ & $4.4^{+0.8}_{-0.9}$ & $0.3^{+0.5}_{-1.8}$\\
\shortstack[l]{\textbf{H3+ data}, \\ Quadrant footprints, without $v_{t,b}$} & $31.2^{+6.4}_{-5.7}$ & $4.6^{+1.7}_{-1.6}$ & $-0.1^{+2.6}_{-3.4}$ & $-30.5^{+5.9}_{-6.6}$ & $10.2^{+2.1}_{-1.9}$ & $4.2^{+0.8}_{-0.7}$ & $-0.2^{+0.7}_{-1.5}$\\
\shortstack[l]{\textbf{H3+ data}, \\ Quadrant footprints, with $v_{t,b}$} & $26.4^{+5.5}_{-4.4}$ & $4.2^{+1.4}_{-1.3}$ & $-1.0^{+2.7}_{-3.6}$ & $-25.3^{+4.5}_{-5.4}$ & $9.2^{+2.3}_{-1.9}$ & $4.4^{+0.7}_{-0.7}$& $-0.1^{+0.7}_{-1.6}$\\
\vspace{-0.3cm}\\
\hline

\end{tabular}
\label{table:posterior-results}
\end{table*}

\subsection{Results in context}\label{sec:results-comparison}

\subsubsection{The reflex motion}

Previously, the reflex motion has been described by the magnitude of the velocity dipole vector, namely the travel velocity, $v_{\rm{travel}}$, and its orientation is called the \textit{apex} direction of the reflex motion $(l_{\rm{apex}}$, $b_{\rm{apex}})$ in Galactic coordinates \citep[e.g.,][]{Vasiliev2021, Petersen2021, Yaaqib2024, Bystrom2025, Chandra2025a}. 
These recent studies agree that the direction of the travel velocity points towards a location on the past orbit of the LMC, though they do not converge on a consistent direction \citep[e.g.,][fig.~9]{Bystrom2025}. 
In this work, instead of transforming to Galactic coordinates, we return the posterior constraints on the Galactocentric Cartesian components of the travel velocity vector i.e., $\vec{v}_{\mathrm{travel}} = (v_x, v_y, v_z)$. 
This is because these are the parameters the MLE fits for in our model to avoid inefficient convergence, see Sec.~\ref{sec:simulations-stellarhaloes}. 
Although, once the posterior has been determined for $\{ v_x, v_y, v_z\}$, one is free to transform to any preferred choice of coordinate system.
The velocity component posterior distributions are shown alongside literature values in Fig.~\ref{fig4}; for a tabulated comparison, see Table.~\ref{table:posterior-results}.
We show the measured median values from \citet[][yellow cross]{Vasiliev2021}, \citet[][SDSS/SEGUE data, $r_{\rm gal}>50 \, \rm kpc$ distance bin, orange diamond]{Yaaqib2024}, \citet[][DESI data, blue circle]{Bystrom2025} and \citet[][H3+SEGUE+MagE data, distance-independent best fit values in their table.~4, red square]{Chandra2025a} in each panel.
We find that for both DESI and H3+SEGUE+MagE data, no matter the chosen on-sky footprint, the $v_x - v_y$ plane preferentially agrees with the measurement of \citet{Vasiliev2021} and the distance-independent measurement of \citet{Chandra2025a}. In fact, the prior distribution (grey contours) struggles to allow values consistent with \citet[][for positive $v_y$]{Yaaqib2024}, \citet[][for negative $v_y$]{Bystrom2025}.
On the other hand, the $v_x - v_z$ plane posteriors tend to agree more with \citet{Bystrom2025}. In that study, the components of the travel velocity are constrained as free parameters. Whereas, in \citet{Vasiliev2021} and this work, they are somewhat constrained by the simulation priors themselves. 
Furthermore, the prior and posteriors in the $v_y - v_z$ plane struggle to be consistent with any of the previous measured values. 
While the posteriors in this work show disagreement with existing measurements in some velocity projections, those measurements are also inconsistent with each other. This further highlights the importance of the intrinsic properties of the datasets themselves e.g., survey selection functions, on the reported measurements on the reflex motion parameters.

\begin{figure*}
    \centering
    \includegraphics[width=0.75\linewidth]{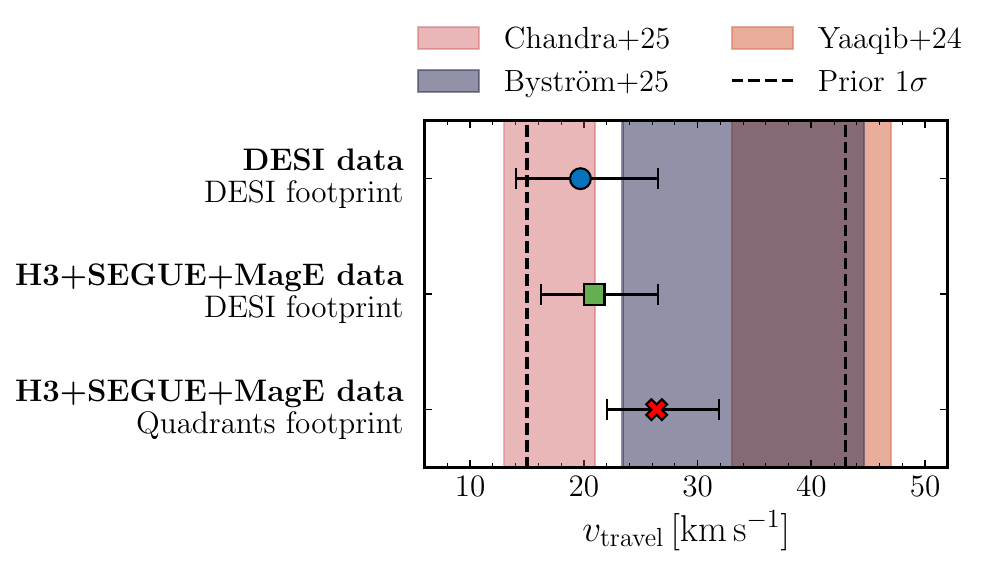}
    \caption{\textbf{Comparison of reflex motion constraints:} The derived medians and $16^{\rm th}-84^{\rm th}$ percentiles of the magnitude of the travel velocity, $v_{\rm travel}$, for the posteriors shown in Fig.~\ref{fig4}. 
    The labels define the dataset used, the on-sky selection footprint used produce the constraint. 
    The values from existing literature are shown as shaded bands \citep[][in orange using SDSS/SEGUE data, red using H3+SEGUE+MagE data, and blue using DESI data, respectively]{Yaaqib2024, Chandra2025a, Bystrom2025}.
    The prior $1\sigma$ confidence interval is shown as the black dashed lines.
    The most precise constraint is produced by using the full depth of the H3+SEGUE+MagE dataset and dividing the binned velocity fields into four quadrants.7}
    \label{fig5}
\end{figure*}

We summarise the magnitude of the travel velocity in Fig.~\ref{fig5}, as derived from the posteriors shown in Fig.~\ref{fig4}. Also see Table.~\ref{table:posterior-results} for tabulated values. 
We present the median and $16^{\rm th}-84^{\rm th}$ percentiles as the points with errors.
The dataset and on-sky footprint used to provide each constraint are shown in the label on the left-hand side. 
As shaded bands, we show the measured $16^{\rm th}-84^{\rm th}$ percentiles confidence intervals from \citet[][best fit values for $r_{\rm gal}>50 \, \rm kpc$ using SDSS/SEGUE data, in orange]{Yaaqib2024}, \citet[][distance-independent best fit values using H3+SEGUE+MagE data, in red]{Chandra2025a} and \citet[][DESI data, in blue]{Bystrom2025}. 
We find that the constraints in this study are smaller in magnitude than those previously measured by \citet{Vasiliev2021} and \citet{Bystrom2025}, but larger in magnitude than the distance-independent measurement of \citet{Chandra2025a}.
This is perhaps unsurprising as we compute the travel velocity in our simulations using all stars beyond $50 \, \rm kpc$, and therefore any constraint on the travel velocity in this work should be interpreted as the mean value in the outer MW halo beyond $50 \, \rm kpc$. 
Hence, the average radial distance of star particles in our stellar haloes is possibly closer/further away than those quoted from the observational datasets. 
Moreover, the medians of the previous literature values are generally larger in magnitude than the median of the prior distribution. 
Hence, the SBI estimates of the travel velocity could be expected to be biased to lower values based purely on the simulation prior alone. 
Finally, we do not consider the effects of radial variation which may bias the constraint on our travel velocity; see Sec.~\ref{sec:discussion-future-models}.
Nevertheless, we obtain our most precise constraint on the travel velocity using the radial and tangential velocity data from H3+SEGUE+MagE in combination with the quadrant sky footprints. Using this dataset, we report a measured travel velocity of $v_{\rm{travel}} = 26.4^{+5.5}_{-4.4} \, \rm km \, \rm s^{-1}$.

\subsubsection{The enclosed mass of the LMC}

A variety of techniques have been used to constrain the enclosed mass of the LMC, e.g, dynamical models of MW stellar streams \citep{Shipp2021, Vasiliev2021, Koposov2023, Warren2025}. 
All of the constraints in this study agree with the previous measurement within $16^{\rm th}-84^{\rm th}$ percentiles uncertainties from e.g., \citet{Koposov2023}. 
The prior of the enclosed LMC mass could be biasing our results to agree with previous measurements. 
Yet, as shown in \citet[][sec.~6.2]{Brooks2026}, adopting a wide uninformative prior instead of an informative Gaussian prior for the LMC mass did not bias our inference of the LMC mass. 
Therefore, we can take confidence that the choice of an informative prior is not significantly biasing the returned constraints in this work.
We obtain our most precise constraint on the enclosed LMC mass using the radial and tangential velocity data from H3+SEGUE+MagE in combination with the quadrant sky footprints. Using this dataset, we report a measured LMC mass enclosed of $M_{\rm LMC}(< 50 \, \rm kpc) = 9.2^{+1.9}_{-2.3} \times 10^{10}\, \rm M_{\odot}$.

\subsubsection{The enclosed mass of the Milky Way}

The MW mass has been constrained using a range of techniques \citep[see,][and the many references therein]{Wang2020, Medina2025}.
The previous measurement from e.g., \citet{Erkal2019}, agrees with all of the various constraints in this work within its $16^{\rm th}-84^{\rm th}$ percentiles confidence intervals.
However, our constraints on the MW enclosed mass are not particularly precise with respect to \citet{Erkal2019}.  
For all constraints, the median and uncertainties are dominated by the enclosed MW mass prior distribution. 
This implies that the mean radial and/or tangential velocity data points are not particularly constraining of the MW mass in this inference set-up. 
Previously, the mass ratio of the MW--LMC system has been suggested to be most sensitive to the velocities of outer halo stars \citep{Petersen2021}.
The apparent model insensitivity to the MW mass implies that the mean velocities of outer halo stars are most informative of the LMC mass and not the mass ratio \citep[see also,][]{Sheng2025, Yaaqib2025}. 
Indeed, future work will use the dispersions of the velocities to constrain the MW mass; see Sec.~\ref{sec:discussion-future-models}. 
Nevertheless, for all of the posteriors in this work conditioned on the various datasets and selection criteria, we find the total mass of the LMC is at least $\approx10-15 \%$ that of the MW \citep[Consistent with][]{Chandra2025a}.
Although, using more generalised models of the MW e.g., triaxiality to define the MW halo density profile, may lead to better constraints on properties of the MW. 

\subsubsection{Dynamical friction}

In our MW--LMC models, we vary the strength of dynamical friction, $\lambda_\mathrm{DF}$, that the LMC experiences. We report the median value and the $16^{\rm th}-84^{\rm th}$ percentile confidence intervals from the posterior distributions in Table.~\ref{table:posterior-results}. 
Regardless of the dataset or sky coverage, we cannot well constrain the dynamical friction strength. 
For the constraints using DESI or H3+SEGUE+MagE data within the DESI footprint, the mean of the posterior distribution is slightly greater than the standard Chandrasekhar value i.e., $\log_{10}(\lambda_{\rm DF}) = 0$.
Interestingly, this is similar to the results in \citet{Koposov2023} where they use measurements of the Orphan-Chenab stellar stream observables to constrain the dynamical friction strength. Their inference is more precise, likely because the Orphan-Chenab stream's trajectory constrains the closest LMC--stream separation distance which depends on the dynamical friction strength.
Conversely, the constraints using the H3+SEGUE+MagE data in on-sky quadrants finds the mean to be slightly smaller than the standard Chandrasekhar values. 
However, the posterior $16^{\rm th}-84^{\rm th}$ percentiles are large and still encompass the standard Chandrasekhar value in all cases.

\begin{figure*}
    \centering
    \includegraphics[width=\linewidth]{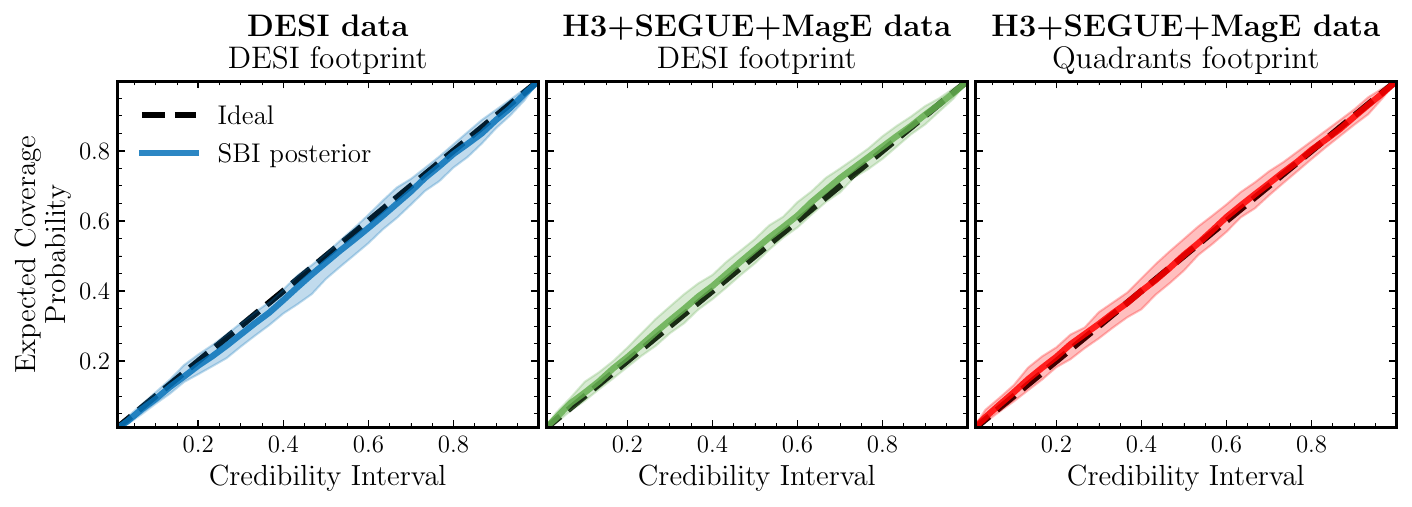}
    \caption{\textbf{Coverage probability posterior check:} For the posteriors estimated using SBI in this work, the probability of finding true test parameters in the appropriate credible intervals matches the expected coverage probability. A bootstrapped $2\sigma$ uncertainty is shown as the shaded region. This validates the estimated SBI posterior distributions and allows one to interpret the confidence intervals on parameter constraints as representative and reliable.}
    \label{fig6}
\end{figure*}

\section{Posterior Diagnostic Checks}\label{sec:posterior-checks}

To ensure the results presented in this work can be trusted, we carry out posterior coverage probability checks and predictive posterior checks.

\subsection{Coverage probability test}

Any posterior from a generative approach should be assessed for its accuracy through a variety of diagnostic tests to gain trust that the inference has been successfully performed.
A coverage probability test is one way to assess the accuracy of such an estimated posterior.
In Bayesian analysis, a coverage test checks whether credible intervals have the expected probabilities \citep[see,][sec.~2.4 for a concise explanation]{Jeffrey2025}. 
In a 1-dimensional posterior setting, one can define a particular credible interval to be
the narrowest interval containing, for example, $90\%$ of the probability weight.
The Bayesian inference procedure takes in some observed data, $D_{\rm obs}$, and determines a posterior distribution, $p(\theta | D_{\rm obs})$, and hence a credible interval for $\theta$. 
For a coverage test, one uses a test parameter, $\theta_{\rm test}$, selected from the prior, $p(\theta)$, as the input to a simulation that produces the corresponding output data point, $D_{\rm test}$. From this, one can derive a posterior, $p(\theta | D_{\rm test})$, and therefore a credible interval. If the inference process has been correctly implemented, then the true test parameter value, $\theta_{\rm test}$, will fall in this credible interval, in this example, $90\%$ of the time. 
Repeating this test for many sampled $\theta_{\rm test}$, and varying the credibility intervals, one can gain trust that the estimated posterior distributions are accurate and have reliable confidence intervals.

To perform a coverage test on our SBI posteriors, we use ‘Tests of Accuracy with Random Points' \citep[TARP,][see their figs.~1\&2 for further intuition]{Lemos2023}.
For our application of SBI, this test is relatively straightforward as we have many pre-existing simulations with an amortised inference scheme, i.e., each data evaluation is computationally cheap without the need to retrain the neural network \citep{Mittal2025}.
TARP coverage probabilities test the accuracy of estimated posteriors by only using samples from the posterior. 
This technique is similar to simulation based calibration \citep{Talts2018} but extends the idea to the full-dimensional posterior space instead of being restricted to 1-dimension. We use the implementation of TARP in the \texttt{sbi} Python package \citep{tejero-cantero2020sbi}.

In Fig.~\ref{fig6}, we demonstrate that the expected coverage does indeed match the credibility level for the estimated posteriors in this work conditioned using the LMC present-day position and velocity, and the radial and tangential velocities of outer halo stars. 
Note, we find similar results for the posteriors conditioned without using the mean tangential data as well. 
This validates our neural posterior estimation as being truly representative of the probability that each of our model parameters has some true value with truly representative uncertainties.
This can be further quantified in two ways. 
Firstly, we can compute the area between the ideal TARP curve and our posterior TARP curves for credibility intervals greater than $0.5$; namely, the Area To Curve (ATC) value. This number should be close to $0$, a value $\gg0$ indicates an estimated posterior that is too wide, conversely, a value $\ll0$ indicates that the estimated posterior is too narrow. 
Secondly, we can calculate the p-value of a Kolmogorov-Smirnov test. The null hypothesis is that an exact one-to-one curve and our posterior TARP curve are identical. If this p-value is less than $0.05$, then this null hypothesis is rejected. 
For all of the estimated posterior distributions, Figs.~\ref{fig:app1}-\ref{fig:app3}, we report an ATC magnitude $\lesssim 0.1$ and a Kolmogorov-Smirnov p-value of $1.0$. 
These values suggest that we are not drastically over-/under-fitting, or biased, and are not required to reject the posterior. 

\subsection{Posterior Predictive Check}

\begin{figure}
    \centering
    \includegraphics[width=\linewidth]{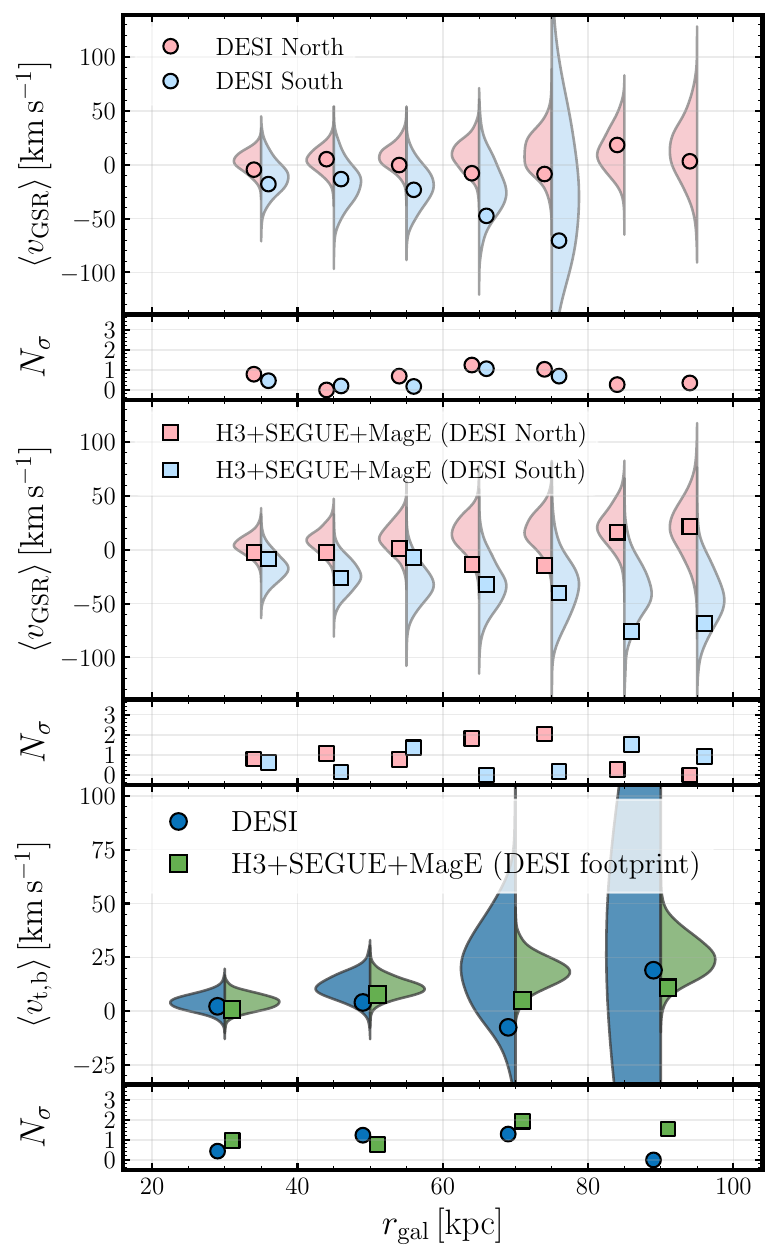}
    \caption{\textbf{Predictive posterior check - DESI / H3+SEGUE+MagE data, DESI footprint:} 
    We test the quality of the estimated posterior distribution returned using the DESI and H3+SEGUE+MagE data within the DESI north and south footprints.
    We show the generated radial (top and middle main panels for DESI and H3+SEGUE+MagE footprints, respectively) and tangential (bottom main panel) velocity data using samples from these posteriors as the violin plot contours. 
    The mean observed data points are shown as the same coloured markers.
    We do not show the data errorbars as the generated data already accounts for the survey uncertainties, allowing direct comparison to mean observed values.
    The sub-panels show the number of standard deviations each observed data point lies from the peak of the generated distribution.
    The generated and original data look sufficiently similar implying the SBI posteriors are representative of the observed data.} 
    \label{fig7}
\end{figure}

\begin{figure}
    \centering
    \includegraphics[width=\linewidth]{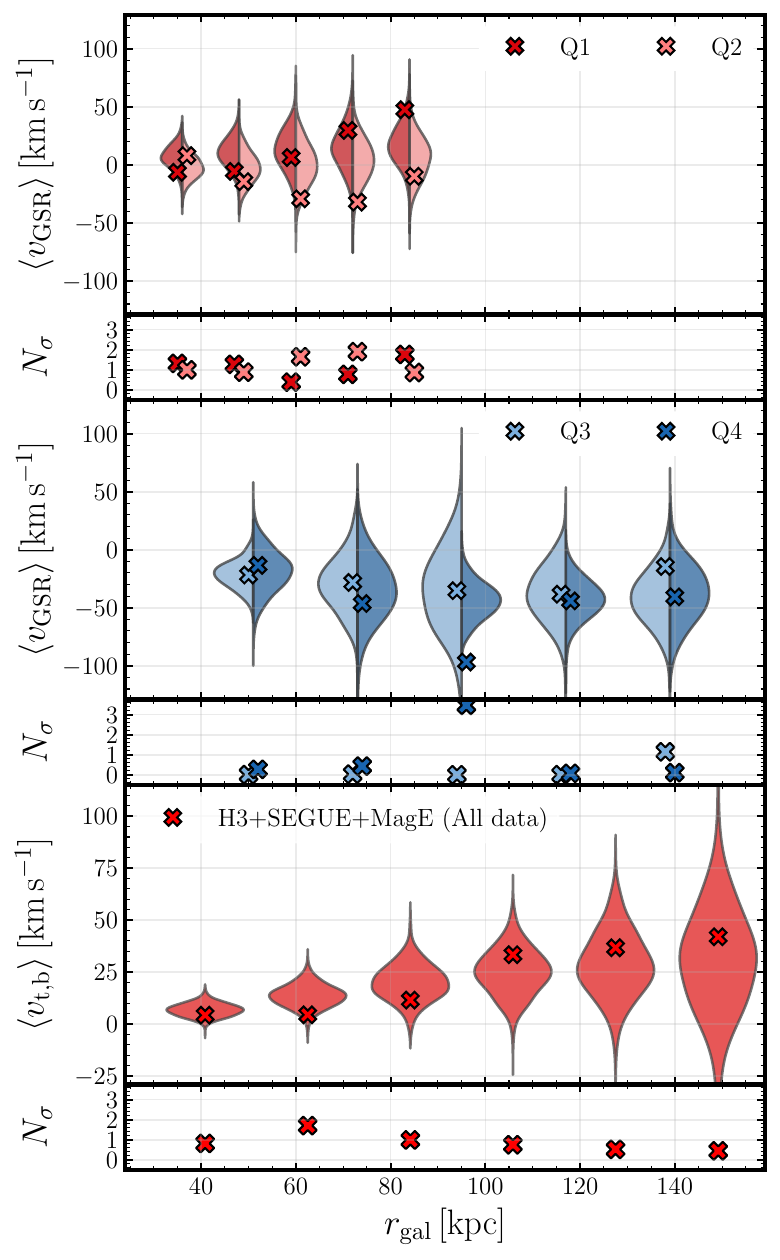}
    \caption{\textbf{Predictive posterior check - H3+SEGUE+MagE data, Quadrants footprints:}
    We test the quality of the estimated posterior distribution returned using H3+SEGUE+MagE data and on-sky quadrant footprints.
    We show the generated radial (top and middle main panels for northern and southern quadrants, respectively) and tangential (bottom main panel) velocity data using samples from this posterior as the violin plot contours. 
    The observed data points are shown as the same coloured crosses.
    We do not show the data errorbars as the generated data already accounts for the survey uncertainties, allowing direct comparison to mean observed values.
    The sub-panels show the number of standard deviations each observed data point lies from the peak of the generated distribution.
    The generated and original data look sufficiently similar implying the SBI posteriors are representative of the observed data.} 
    \label{fig8}
\end{figure}

We carry out a ‘Posterior Predictive Check' (PPC) to act as a complementary diagnostic test. This test makes use of the fact that if the inference has been correctly implemented, then any generated data, $D_{\rm pp}$, using simulation parameters as sampled from the posterior, $\theta_{\rm pp}$, should be similar to the observed data, $D_{\rm obs}$ \citep[][]{Lueckmann21a}. A PPC can provide an intuition about any bias introduced in inference e.g., determining whether or not the generated data systematically differ from the observed data used during the inference.
We carry out a PPC for every posterior used throughout this work, see Figs.~\ref{fig:app1}-\ref{fig:app3}.
To do this, we first sample model parameter values from these posterior distributions. 
We then re-run rigid MW--LMC simulations adopting these parameter values, generating output data as the average radial and tangential velocity measurements, which we can then use to compare to the observed data that was originally used to evaluate the posterior. 

In Fig.~\ref{fig7}, we demonstrate the PCC for the posteriors conditioned on DESI or H3+SEGUE+MagE data within the DESI survey footprint, i.e., open blue and green contours in Figs.~\ref{fig3} \& \ref{fig4}.  
Using parameters sampled from these posteriors, we show the generated radial (top and middle panels for DESI and H3+SEGUE+MagE footprints, respectively) and tangential (bottom panel) velocity data using samples from these posteriors as the violin plot contours. 
The mean observed data points are shown as the same coloured markers; same points as in Fig.~\ref{fig2}.
Errorbars are not shown on the observed data because the generated data incorporates survey uncertainties. 
The sub-panels in Fig.~\ref{fig7} demonstrate the number of standard deviations that each observed data point lies away from the peak of the generated data distribution.
Mostly, the observed data points lie within $\sim 2 \sigma$ the generated data distributions, this demonstrates that the SBI posteriors are accurately representing the observations.

In Fig.~\ref{fig8}, we demonstrate the PCC for the posteriors conditioned on the full H3+SEGUE+MagE dataset with on-sky quadrant footprints, i.e,. open red contours in Figs.~\ref{fig3} \& \ref{fig4}.
Using parameters sampled from this posterior, we show the generated data for the average radial (top and middle panels, for quadrants 1 \& 2 and 3 \& 4, respectively) and tangential (bottom) velocity data as the violin plot contours.
The mean observed data points are shown as the same coloured markers; same points as in Figs.~\ref{fig2}.
Again, errorbars are not shown because the generated data incorporates survey uncertainties.
In most distance bins the observed data is found to be truly represented by data generated from posterior sampling, i.e. within $\sim 2 \sigma$ as seen in the sub-panels.
However, the observed data for the northern quadrants, Q1 and Q2, is less well represented by the generated data than the southern quadrants, Q3 and Q4. 
This is likely a limitation using this suite of reasonably simplistic rigid MW--LMC simulations which are able to capture global velocity perturbations, e.g. dipole signatures, but struggle to capture smaller scale perturbations e.g., quadrupole signatures.
However, even higher fidelity models of the MW--LMC system, e.g., deforming simulations \citep{Garavito-Camargo2019}, struggle to reproduce these velocity trends in the northern Galactic hemisphere \citep[see fig.~4,][]{Chandra2025a}.
Additionally, the observed data point in Q4 around $80-100 \, \rm kpc$ exhibits behaviour that is inconsistent with the simulations; $\sim 3 \sigma$ discrepant from the peak of the generated distribution.
However, this has been seen previously in \citet{Chandra2025a} and been suggested to be due to unresolved substructure at these distances, for example stars stripped from the LMC. 
Finally, given the current precision of the observations, the rigid models and generated data in this study sufficiently represent the observations; see Fig.~\ref{fig7} \& Fig.~\ref{fig8}. In the future, as the observational precision increases, one must ensure this remains true otherwise more complex models and simulations may be required to operate within an SBI framework.

\section{Discussion}\label{sec:discussion}

\subsection{Limitations and caveats}

In \citet[][sec.~6.1]{Brooks2026} we discussed many of the limitations and caveats of using rigid models for this analysis. These include: assuming that the MW is at equilibrium prior to the LMC's infall; that the LMC is on its first pericentric passage; that there is no mass exchange of the MW and LMC; that the MW stellar haloes are non-triaxial and that dynamical friction is sufficiently well captured using a Chandrasekhar prescription. We direct the reader to that section for suggestions on how the model can be improved to account for those limitations. 
We also note that using rigid models fixes the MW disc at the centre of the Galaxy. Hence, the reflex motion cannot discern between the motion of the disc relative to the outer MW halo and the internal bulk motions of the MW stellar halo induced by the LMC \citep{Yaaqib2025}. 
Extending the MW--LMC models to include self-gravity, e.g., using the $N$-body simulations in \citet{Sheng2025} to train the SBI neural networks, could in theory provide a way to navigate this effect and further improve the system modelling by recovering the full dipole signal \citep{Petersen2020}. 
Nevertheless, as shown in \citet{Brooks2026}, training an SBI framework on many rigid simulations retains enough of the relevant physics that more complex simulations capture \citep[e.g., deforming N-body simulations,][]{Garavito-Camargo2019} to avoid model misspecification and allow rapid exploration of large model parameter spaces at a fraction of the computational cost required to run the higher fidelity simulations.

In this study, one of the limitations is the choice of summary statistics used for the parameter inference. 
As a reliable and simple starting point, we implemented a binned radial and tangential velocity field approach, see Sec.~\ref{sec:data-measurements}. By computing average values for the radial and tangential velocities in distance bins this leads to data compression over such scales, thus potentially missing out on some constraining power. 
The most powerful way to constrain the MW, LMC and reflex motion parameters would be to employ a star-by-star fitting procedure. 
However, as there are $\mathcal{O}(10^3)$ stars in any given dataset, this would mean the dimensionality of the inference framework would increase dramatically, potentially subjecting this methodology to the curse of dimensionality. 
Another approach to improve upon the current set of summary statistics would be to increase the number of distance bins and reduce the area on the sky over which the binned velocities are computed. Although, increasing the number of bins would require one to ensure the model can still reliably produce realistic observed data points.
This naturally lends itself to implement a spherical harmonic expansion to the halo velocity fields in a set appropriately defined distance bins \citep[e.g,][]{Cunningham2020}. 

In any approach, the selection function of a survey should be accounted for. In this work, we consider the depth and on-sky footprint of a survey, but do not account for any non-uniformity, for example the grouping of sources near the survey footprint boundaries in DESI. 
As an intermediate step, a nearest neighbours algorithm could be used to select the same number density of observed sources as a function of distance and angle on-sky. 
This would reduce the risk of introducing any systematic biases during the inference as the accuracy of the forward modelled systems has been improved by accounting for another observational effect. 

Finally the choice of prior distribution for, e.g., the MW and LMC masses, influences the resulting inference. 
In \citet[][sec 6.2]{Brooks2026}, we found that using both informative and uninformative priors produced similar MW and LMC mass posteriors.
Within this study, we find the MW mass posterior is insensitive to the provided data, and the inference is prior dominated.
Further to this, the maximum probability for the LMC mass prior and posterior are similar, although the latter is more precise; see Fig.~\ref{fig3}. 
That being said, these posteriors are not in tension with the prior as the observed data is well represented by the use of rigid MW--LMC models; see the PPC in Fig.~\ref{fig7} \& Fig.~\ref{fig8}.
Nevertheless, we note that the presented results cannot be taken as an entirely prior-independent measurements.

\subsection{Sky coverage}\label{sec:discussion-skycoverage}

We have varied the area over which the binned velocities were calculated by either using the DESI northern and southern survey fields or defined quadrants. 
In principle, as long as there is sufficient good-quality data, using the quadrants with a finer on-sky division of the data is expected to produce results that are more accurate and precise simply because of averaging over smaller sky coverages.  
Using the all-sky H3+SEGUE+MagE survey data, we can qualitatively assess the importance of sky coverage on returned model parameter constraints. 
For example, in Fig.~\ref{fig5}, we can compare the travel velocity values found using the H3+SEGUE+MagE and choice of sky coverage. 
The use of the quadrants sky coverage leads to a larger value than the DESI fields, although within the uncertainties they agree with each other. 
Therefore, there appears to be a subtle degree of sensitivity to returned constraints given the imposed on-sky selection. Current and upcoming all-sky datasets are best placed to further investigate this effect.

\subsection{Data dimensionality}\label{sec:discussion-tangentialvels}

We have varied whether we include information on the reflex corrected tangential velocities as the input data used to set parameter constraints. 
In general, including the tangential velocities slightly improves the inference precision. However, the effect is almost negligible for DESI and only just noticeable for H3+SEGUE+MagE. 
This can be understood from the measured uncertainties from DESI and H3+SEGUE+MagE in Sec.~\ref{sec:data-measurements-vt} and the panel (e) of Fig.~\ref{fig2}. As the measured uncertainties are large, $\mathcal{O}(10\,\rm km \,\rm s^{-1})$, when we produce survey-specific measurements from our simulations that are used to train the neural posterior estimator, this allows for a wide range of models that can explain the observed mean data points.
As more precise distance and proper motion measurements become available with the release of upcoming datasets e.g., RR Lyrae stars in Gaia Data Release 4 (DR4), this will reduce the uncertainties in the measured binned tangential velocities and improve the precision on the inferred model parameters. Plus, the increase in the number of stars with available proper motion measurements e.g., Sloan Digital Sky Survey V \citep[SDSS-V,][]{SDSSCollaboration2025}, will improve inference as well.

Throughout our inference, we have only used the mean radial and tangential velocities of outer halo stars. In future work, we can extended the current inference set-up to use information on the velocity dispersions as well; see Sec.~\ref{sec:discussion-future-models}. 

We assumed knowledge of the present-day LMC position and velocity. 
There remains ambiguity in the definition of the LMC centre \citep[see sec.~4.1,][]{vanderMarel2014} which in turn corresponds to different values of the mean proper motions \citep{GaiaCollaboration2018, GaiaCollaboration2021, Wan2020}. 
We use well motivated LMC coordinates as input data points such that posterior probability weight is not unfairly pushed to include unphysical solutions. This helps to break model degeneracies e.g., for the dynamical friction strength. That being said, the data which provides the most constraining power on model parameters is the velocities of the outer halo stars.
An interesting future application of this SBI architecture would be to constrain the present-day LMC centre's position and velocity using the dynamics of outer halo stars as tracers. 

\subsection{Extending the Milky Way--LMC models}\label{sec:discussion-future-models}

\noindent
There are some improvements to the current models and methodology which will benefit future studies using this SBI approach.
Firstly, we will account for radial variation in the reflex motion by implementing a linear continuity model \citep{Chandra2025a}. 
As the amplitude of the reflex motion is expected to increase with Galactocentric distance \citep{Yaaqib2024,Yaaqib2025} this will allow the simulation priors to cover a larger area of this parameter space and allow for a more fair comparison to previous literature results.
Secondly, we have only considered using the \textit{mean} of the radial and tangential velocities  of outer halo stars in this work to perform the inference. 
Future analysis, see \citet{Brooks2026b}, will include the second-order velocity moments, i.e., the velocity \textit{dispersions}, of outer halo stars as this allows constraints to be placed on the MW mass \citep{Jeans1915, Binney2008, Sheng2025} and the velocity anisotropy of the stellar halo \citep[e.g.,][]{Cunningham2019, Bird2021, Han2024, Chandra2025a}.
Finally, we assumed the MW to be in equilibrium prior to the LMC's infall. This model assumption is not strictly true as there has been observed net compression \citep[e.g.,][]{Chandra2025a}, and net rotation \citep[e.g.,][though alone is not a disequilibrium state]{Deason2017}, of the MW's stellar halo. 
Accounting for the pre-processed state of the MW's stellar halo prior to the LMC's infall is essential for future studies aiming to constrain the LMC's past orbit \citep{Sheng2025}. Moreover, accounting for these effects may help in generating simulated data that better represents all observed data points.

\section{Conclusions \& Outlook}\label{sec:conclusions}

We have presented an SBI architecture to infer MW--LMC parameters using the velocities of outer halo stars and provided its first application the DESI and H3+SEGUE+MagE survey datasets. 
The SBI framework is trained on a large set of $128,000$ rigid MW--LMC simulations, with each stellar halo containing $20,000$ star particles.
We account for the survey-specific uncertainties and use the average radial and tangential velocities of these stars as a function of distance to constrain model parameters. 
We summarise our key findings as follows:

\begin{enumerate}
    \item We obtain our most precise MW--LMC parameter constraints using the average radial and tangential velocities of outer halo stars, divided into on-sky quadrants, from the all-sky H3+SEGUE+MagE dataset: 
        \begin{description}
            \item[\textbf{The reflex motion velocity}, $v_{\rm{travel}} = 26.4^{+5.5}_{-4.4} \, \rm km \, \rm s^{-1}$] 
            \vspace{0.25cm}
            \item[\textbf{The enclosed LMC mass}, $M_{\rm LMC}(< 50 \, \rm kpc) = 9.2^{+1.9}_{-2.3} \times 10^{10}\, \rm M_{\odot}$] 
            \vspace{0.25cm}
        \end{description}
    \noindent
    We note the precision in the distance-averaged reflex motion velocity is slightly larger than previously found in \citet{Chandra2025a} using the same dataset but unique methodology.
    \item We find that the infall LMC's total mass is at least $\approx10-15 \%$ of the MW's total mass. This result is consistent across the DESI and H3+SEGUE+MagE datasets, independent of the velocity information or on-sky footprint selection used.
    \item Given the prior distributions, we find that the returned MW mass is insensitive to the mean radial and tangential velocities of outer halo stars. Future use of the dispersions of these velocity components will help to break model degeneracies \citep{Brooks2026b}.
    \item For both the DESI and H3+SEGUE+MagE datasets, when using the DESI sky coverage, constraints on $v_{\rm{travel}}$ and $M_{\rm LMC}$ are systematically pushed towards lower values relative to using on-sky quadrants, but remain consistent within uncertainties.
    \item The inclusion of tangential velocities of outer halo stars as input data for the inference process provides minimal extra constraining power for MW--LMC model parameters. This is because the measured tangential velocity uncertainties remain large. Upcoming data releases e.g., Gaia DR4, will greatly improve their precision and in turn the constraints on MW--LMC model parameters.
    \item We find little preference for the strength of dynamical friction, $\lambda_{\rm DF}$, to deviate from the standard Chandrasekhar value. 
    \item We have developed an SBI framework that enables rapid inference across large MW--LMC parameter spaces while incorporating time dependence. This approach will be essential for exploiting upcoming outer halo datasets.
\end{enumerate}

\noindent
We look forward to the upcoming data releases from SDSS-V and Gaia DR4 to provide an increased number of outer halo stars with precise velocity measurements.  
Using this exquisite data in combination with the SBI architecture presented in this work, we will produce the most up-to-date, rapid and reliable constraints on the masses of the MW, LMC and the induced reflex motion.

\section*{Acknowledgements}

We thank the referee for their feedback that improved the quality of the manuscript. RANB acknowledges support from the Royal Society and the Flatiron Institute, Simons Foundation. RANB would like to thank all of the H3 team for access to their data.
JLS acknowledges the support of the Royal Society (URF\textbackslash R1\textbackslash191555; URF\textbackslash R\textbackslash 241030). 
NGC acknowledges support from the Heising-Simons Foundation grant \#2022-3927, through which NGC is supported by the Barbara Pichardo Future Faculty Fellowship.

\section*{Data Availability}\label{sec:data-availability}

The DESI DR1 BHB catalogue is available here:~\url{https://data.desi.lbl.gov/doc/releases/dr1/vac/mws-bhb/}. 
The MW--LMC simulation parameters and the data for $128,000$ unique stellar haloes, each with $20,000$ particles, will be shared upon reasonable request. 
The posteriors produced using SBI in this work are available \href{https://zenodo.org/records/17226585}{here}.

\textit{Software:} \texttt{sbi} \citep{tejero-cantero2020sbi}, \texttt{agama} \citep{2019MNRAS.482.1525V}, \texttt{gala} \citep{gala}, NumPy \citep{harris2020array}, Matplotlib \citep{Hunter:2007}, Seaborn \citep{Waskom2021}, \texttt{corner} \citep{corner}, Astropy \citep{astropy:2013, astropy:2018, astropy:2022}, SciPy \citep{2020SciPy-NMeth}.



\bibliographystyle{mnras}
\bibliography{biblio} 




\appendix

\section{Full Posterior Distributions}\label{sec:appendixA}

\begin{figure*}
    \centering
    \begin{overpic}[width=1\linewidth]{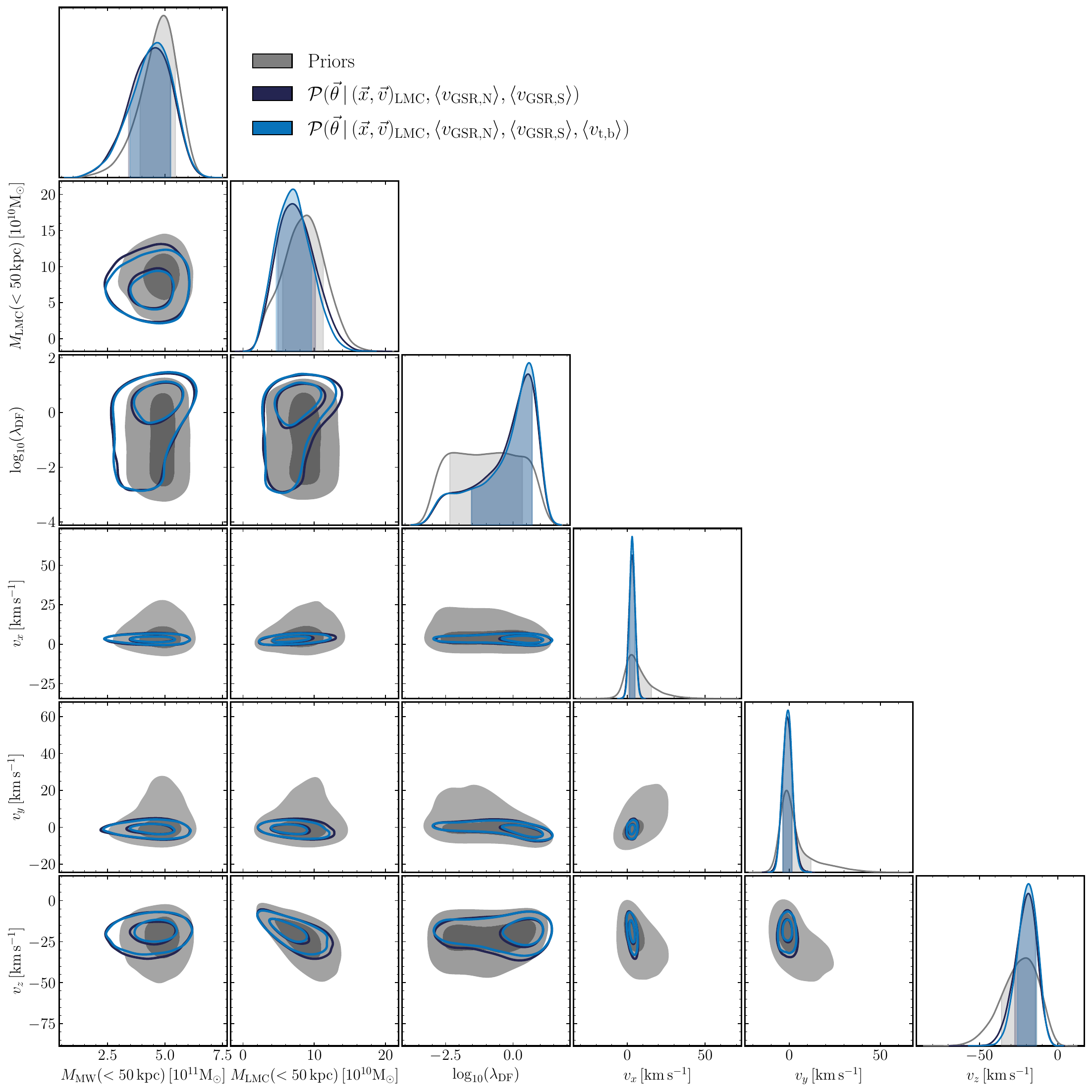}
         \put(57,63){\includegraphics[width=0.44\linewidth]{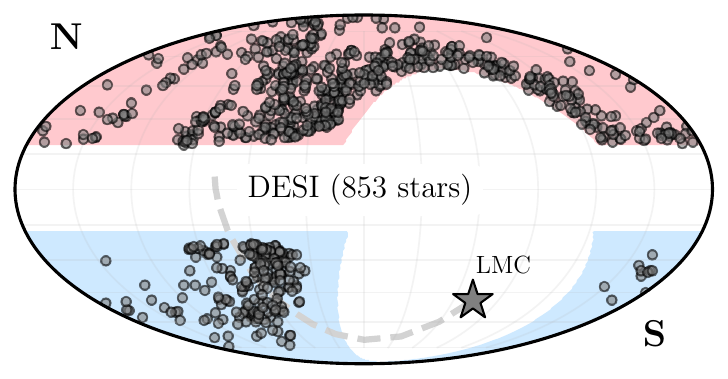}}  
    \end{overpic}
    \caption{\textbf{Posterior distributions - DESI data, DESI footprint:} The joint, and individual, posterior distributions for the MW and LMC masses enclosed within $50\,\rm kpc$, the dynamical friction strength, $\lambda_\mathrm{DF}$, and the Cartesian components of the reflex motion travel velocity. 
    The dark blue and light blue open contours represent the posteriors conditioned without and with the tangential velocity information, respectively.
    The prior distributions are shown as the filled grey contours.
    For the 1D posterior panels we show the $16^{\rm th}-84^{\rm th}$ percentiles as shaded regions. 
    The inclusion of DESI tangential velocity information does not greatly improve the precision of the inferred parameters.
    The selected DESI data between $30-100\,\rm kpc$ and contained within the northern (pink) and southern (blue) DESI survey footprints are shown in the top-right inset.} 
    \label{fig:app1}
\end{figure*}

\begin{figure*}
    \centering
    \begin{overpic}[width=1\linewidth]{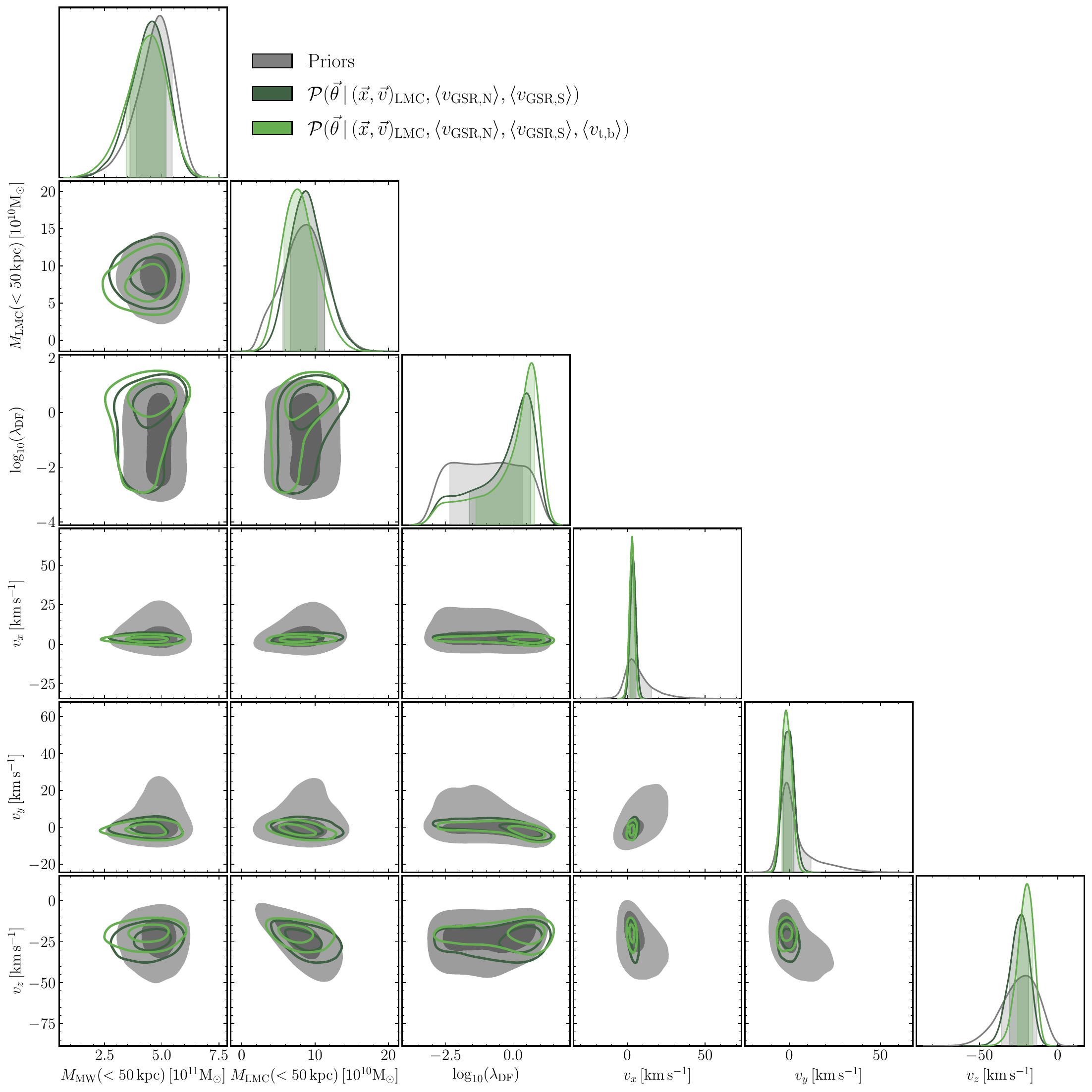}
         \put(57,63){\includegraphics[width=0.44\linewidth]{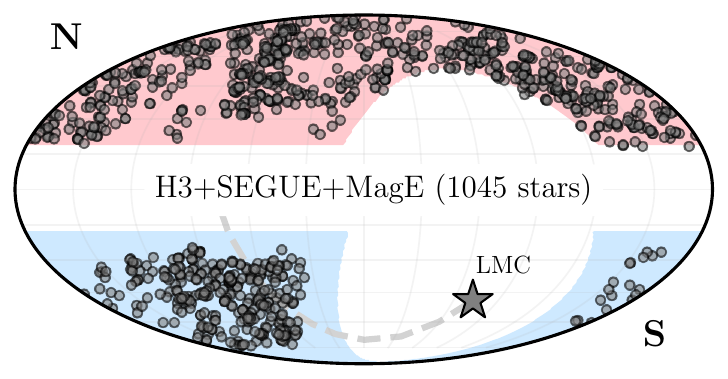}}  
      \end{overpic}
    \caption{\textbf{Posterior distributions - H3+SEGUE+MagE data, DESI footprint:} The joint, and individual, posterior distributions for the MW and LMC masses enclosed within $50\,\rm kpc$, the dynamical friction strength, and the Cartesian components of the reflex motion travel velocity. 
    The dark green and light green open contours represent the posteriors conditioned without and with the tangential velocity information, respectively.
    The prior distributions are shown as the filled grey contours.
    For the 1D posterior panels we show the $16^{\rm th}-84^{\rm th}$ percentiles as a shaded region. 
    The inclusion of the tangential velocity information slightly improves the precision of the inferred parameters.
    The selected H3+SEGUE+MagE data between $30-100\,\rm kpc$ and contained within the northern (red) and southern (blue) DESI survey footprints are shown in the top-right inset.} 
    \label{fig:app2}
\end{figure*}

\begin{figure*}
    \centering
    \begin{overpic}[width=1\linewidth]{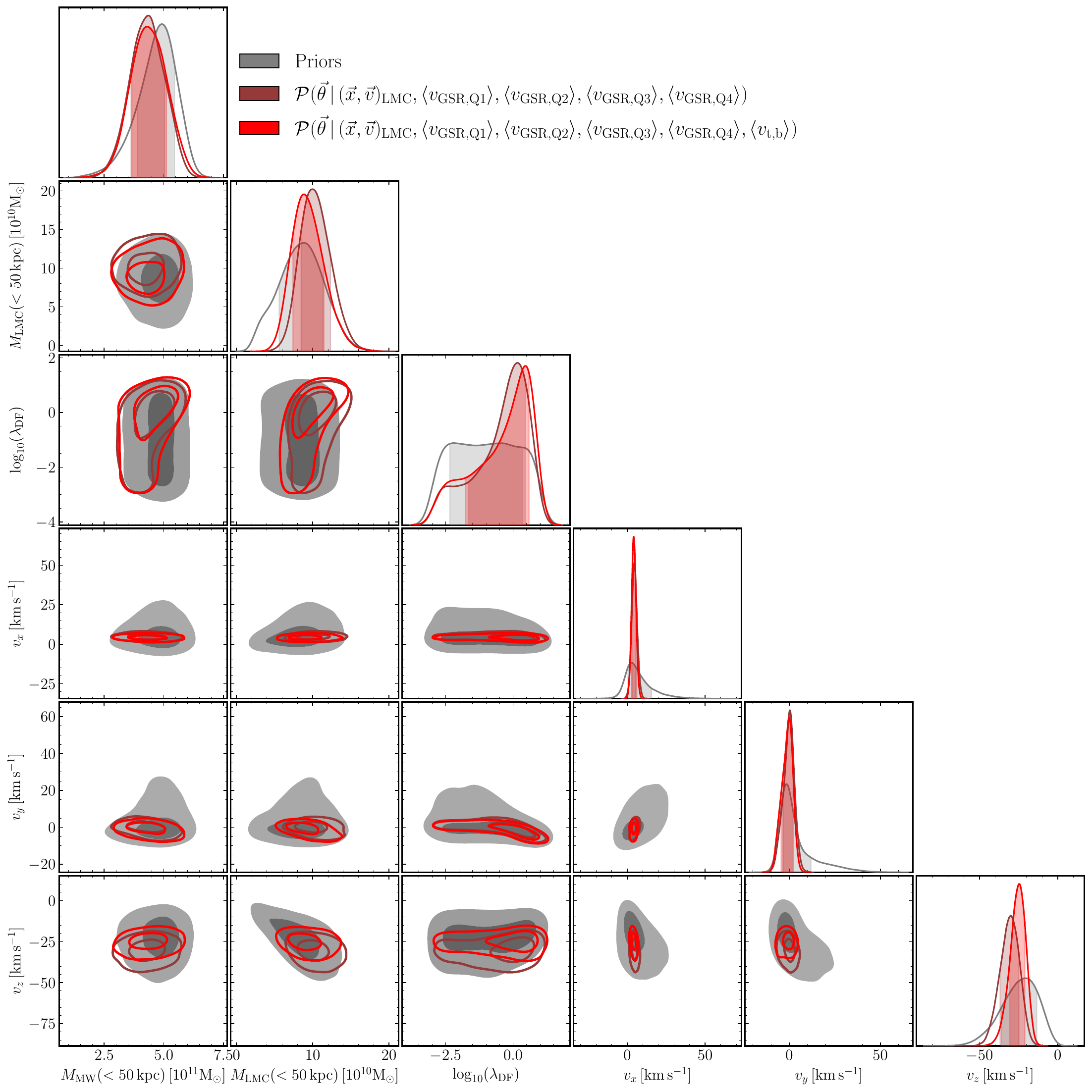}
         \put(57,61){\includegraphics[width=0.44\linewidth]{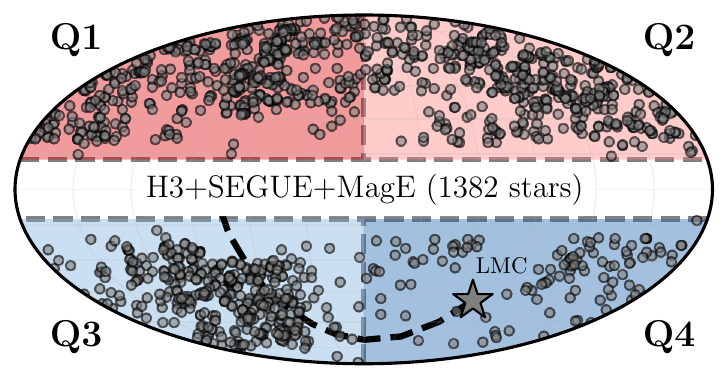}}  
      \end{overpic}
    \caption{\textbf{Posterior distributions - H3+SEGUE+MagE data, Quadrants footprint:} The joint, and individual, posterior distributions for MW and LMC masses enclosed within $50\,\rm kpc$, the dynamical friction strength, and the cartesian components of the reflex motion travel velocity.
    The dark red and light red open contours represent the posteriors conditioned without and with the tangential velocity information, respectively.
    The prior distributions are shown as the filled grey contours.
    For the 1D posterior panels we show the $16^{\rm th}-84^{\rm th}$ percentiles as a shaded region. 
    The inclusion of the tangential velocity information improves the precision of the inferred parameters.
    The selected H3+SEGUE+MagE data between $30-160\,\rm kpc$ and contained within the four defined quadrants, Q1-4, is shown in the top-right inset.} 
    \label{fig:app3}
\end{figure*}

We provide the full posterior distributions for MW and LMC enclosed masses, the strength of dynamical friction and the Galactocentric Cartesian travel velocity components conditioned on different data subsets.
In Fig.~\ref{fig:app1}, we show the posterior conditioned on DESI data. The dark blue and light blue open contours represent the posteriors conditioned without and with the tangential velocity information, respectively; see Sec.~\ref{sec:results-desifoot-desidata}.
In Fig.~\ref{fig:app2}, we show the posterior conditioned on H3+SEGUE+MagE data within the DESI survey footprint. The dark green and light green open contours represent the posteriors conditioned without and with the tangential velocity information, respectively; see Sec.~\ref{sec:results-desifoot-h3data}.
In Fig.~\ref{fig:app3}, we show the posterior conditioned on the full H3+SEGUE+MagE dataset, divided into on-sky quadrants. The dark red and light red open contours represent the posteriors conditioned without and with the tangential velocity information, respectively; see Sec.~\ref{sec:results-quadfoot}.
In all figures, the prior distributions are shown as the filled grey contours. 
Additionally, we show the data used to produce the constraints on the all-sky projection. An illustrative LMC orbit is shown as the dashed line with its present day position denoted by the grey star marker.

\bsp	
\label{lastpage}
\end{document}